%% file: 0_main.tex
\documentclass{sig-alternate-2013}
\usepackage[english]{babel}
\usepackage{tabularx}
\usepackage{graphicx}
\usepackage{url}
\usepackage{amsmath}
\usepackage{mdwlist}
\usepackage{subcaption}
\usepackage{hyperref}
\usepackage[labelfont=bf,textfont=bf]{caption}
\usepackage{multirow}
\usepackage{array}
\usepackage[linesnumbered,ruled]{algorithm2e}
\SetKwComment{Comment}{$\triangleright$\ }{}
\usepackage{xcolor}
\usepackage{epstopdf}

\usepackage[utf8]{inputenc}
\usepackage{amssymb}

\newcommand{\ie}{\emph{i.e., }}
\newcommand{\eg}{\emph{e.g., }}
\newcommand{\etal}{\emph{et al.}}

\newcommand{\wrt}{\emph{w.r.t. }}
\newcommand{\cf}{\emph{cf. }}

\begin{document}
	
\permission{\copyright 2017 International World Wide Web Conference Committee \\ (IW3C2), published under Creative Commons CC BY 4.0 License.}
\conferenceinfo{WWW 2017,}{April 3--7, 2017, Perth, Australia.}
\copyrightetc{ACM \the\acmcopyr}
\crdata{978-1-4503-4913-0/17/04. \\
http://dx.doi.org/10.1145/3038912.3052569 \\
\includegraphics{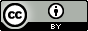}}
	
\clubpenalty=10000
\widowpenalty=10000

\title{Neural Collaborative Filtering\small{\thanks{NExT research is supported by the National Research Foundation, Prime Minister's Office, Singapore under its IRC@SG Funding Initiative.}}}

\numberofauthors{6}

\author{
\alignauthor
Xiangnan He\\
\affaddr{National University of Singapore, Singapore}\\
\email{xiangnanhe@gmail.com}
\alignauthor
Lizi Liao\\
\affaddr{National University of Singapore, Singapore}\\
\email{liaolizi.llz@gmail.com}
\alignauthor Hanwang Zhang\\
\affaddr{Columbia University}\\
\affaddr{USA}\\
\email{hanwangzhang@gmail.com}
\and  
\alignauthor Liqiang Nie\\
\affaddr{Shandong University}\\
\affaddr{China}
\email{nieliqiang@gmail.com}
\alignauthor Xia Hu\\
\affaddr{Texas A\&M University}\\
\affaddr{USA}\\
\email{hu@cse.tamu.edu}
\alignauthor Tat-Seng Chua\\
\affaddr{National University of Singapore, Singapore}\\
\email{dcscts@nus.edu.sg}
}

\maketitle
\begin{abstract}	
In recent years, deep neural networks have yielded immense success on speech recognition, computer vision and natural language processing. 
However, the exploration of deep neural networks on recommender systems has received relatively less scrutiny.
In this work, we strive to develop techniques based on neural networks to tackle the key problem in recommendation --- collaborative filtering --- on the basis of implicit feedback. 

Although some recent work has employed deep learning for recommendation, they primarily used it to model auxiliary information, such as textual descriptions of items and acoustic features of musics. When it comes to model the key factor in collaborative filtering --- the interaction between user and item features, they still resorted to matrix factorization and applied an inner product on the latent features of users and items. 

By replacing the inner product with a neural architecture that can learn an arbitrary function from data, we present a general framework named NCF, short for \textit{Neural network-based Collaborative Filtering}. 
NCF is generic and can express and generalize matrix factorization under its framework. 
To supercharge NCF modelling with non-linearities, we propose to leverage a multi-layer perceptron 
to learn the user--item interaction function. 
Extensive experiments on two real-world datasets show significant improvements of our proposed NCF framework over the state-of-the-art methods.
Empirical evidence shows that using deeper layers of neural networks offers better recommendation performance. 
\end{abstract}
\vspace{-5pt}


\input{1_introduction}
\input{2_preliminary} 
\input{3_method}
\input{4_experiment}
\input{5_related}
\input{6_conclusion}

\bibliographystyle{abbrv}
\small{\bibliography{proc}}
\end{document}

%% file: 1_introduction.tex
\section{Introduction}
\label{sec:introduction}
In the era of information explosion, recommender systems play a pivotal role in alleviating information overload, having been widely adopted by many online services, including E-commerce, online news and social media sites. The key to a personalized recommender system is in modelling users' preference on items based on their past interactions~(\eg ratings and clicks), known as collaborative filtering~\cite{ItemCF,DiscreteCF}. Among the various collaborative filtering techniques, matrix factorization~(MF)~\cite{fastMF,SVD++} is the most popular one, which projects users and items into a shared latent space, using a vector of latent features to represent a user or an item. Thereafter a user's interaction on an item is modelled as the inner product of their latent vectors. 

Popularized by the Netflix Prize,
MF has become the \textit{de facto} approach to latent factor model-based recommendation. Much research effort has been devoted to enhancing MF, such as 
integrating it with neighbor-based models~\cite{SVD++}, combining it with topic models of item content~\cite{Wang:KDD15}, and extending it to factorization machines~\cite{FM} for a generic modelling of features. 
Despite the effectiveness of MF for collaborative filtering, it is well-known that its performance can be hindered by the simple choice of the interaction function --- inner product. For example, for the task of rating prediction on explicit feedback, it is well known that 
the performance of the MF model
can be improved by incorporating user and item bias terms into the interaction function\footnote{\small{\url{http://alex.smola.org/teaching/berkeley2012/slides/8_Recommender.pdf}}}. While it seems to be just a trivial tweak for the inner product operator~\cite{fastMF}, it points to the positive effect of designing a better, dedicated interaction function for modelling the latent feature interactions between users and items. 
The inner product, which simply combines the multiplication of latent features linearly, may not be sufficient to capture the complex structure of user interaction data. 

This paper explores the use of deep neural networks for learning the interaction function 
from data, rather than a handcraft that has been done by many previous work~\cite{Hu:2014,SVD++}. The neural network has been proven to be capable of approximating any continuous function~\cite{Hornik:1989},
and more recently deep neural networks~(DNNs) have been found to be effective in several domains, ranging from computer vision, speech recognition, to text processing~\cite{Collobert:2008,cvpr16best,Hong_group,zhang2014start}.
However, there is relatively little work on employing DNNs for recommendation in contrast to the vast amount of literature on MF methods. Although some recent advances~\cite{van2013deep,Wang:KDD15,Zhang:2016:CKB} have applied DNNs to recommendation tasks and shown promising results, they mostly used DNNs to model auxiliary information, such as textual description of items, audio features of musics, and visual content of images.
With regards to modelling the key collaborative filtering effect, they still resorted to MF, combining user and item latent features using an inner product. 

This work addresses the aforementioned research problems by formalizing a neural network modelling approach for collaborative filtering. We focus on implicit feedback, which indirectly reflects users' preference through behaviours like watching videos, purchasing products and clicking items. Compared to explicit feedback~(\ie ratings and reviews), implicit
feedback can be tracked automatically and is thus much easier to collect for content providers. However, it is more challenging to utilize, since user satisfaction is not observed and there is a natural scarcity of negative feedback. In this paper, we explore the central theme of how to utilize DNNs to model noisy implicit feedback signals. 

The main contributions of this work are as follows.\vspace{-5pt}
\begin{enumerate}
\item We present a neural network architecture to model latent features of users and items and devise a general framework NCF for collaborative filtering based on neural networks. \vspace{-5pt}
\item We show that MF can be interpreted as a specialization of NCF and utilize a multi-layer perceptron to endow NCF modelling with a high level of non-linearities.\vspace{-5pt}
\item We perform extensive experiments on two real-world datasets to demonstrate the effectiveness of our NCF approaches and the promise of deep learning for collaborative filtering. \vspace{-3pt}
\end{enumerate}




%% file: 2_preliminary.tex
\section{Preliminaries}
\label{sec:preliminaries}
We first formalize the problem and discuss existing solutions for collaborative filtering with implicit feedback. We then shortly recapitulate the widely used MF model, highlighting its limitation caused by using an inner product. 

\subsection{Learning from Implicit Data}
Let $M$ and $N$ denote the number of users and items, respectively. We define the user--item interaction matrix $\textbf{Y}\in\mathbb{R}^{M\times N}$ from users' implicit feedback as, \vspace{-5pt}
\begin{equation}
	y_{ui} = 
	\begin{cases} 
	1 , & \text{if interaction (user $u$, item $i$) is observed;}\\
	0 , & \text{otherwise.}
	\end{cases}
\end{equation}
Here a value of 1 for $y_{ui}$ indicates that there is an interaction between user $u$ and item $i$; however, it does not mean $u$ actually likes $i$.
Similarly, a value of 0 does not necessarily mean $u$ does not like $i$, it can be that the user is not aware of the item. 
This poses challenges in learning from implicit data, since it provides only noisy signals about users' preference. While observed entries at least reflect users' interest on items, the unobserved entries can be just missing data and there is a natural scarcity of negative feedback. 
 
The recommendation problem with implicit feedback is formulated as the problem of estimating the scores of unobserved entries in $\textbf{Y}$, which are used for ranking the items. 
Model-based approaches assume that data can be generated (or described) by an underlying model. Formally, they can be abstracted as learning 
$\hat{y}_{ui} = f(u, i | \Theta),$
where $\hat{y}_{ui}$ denotes the predicted score of interaction $y_{ui}$, $\Theta$ denotes model parameters, and $f$ denotes the function that maps model parameters to the predicted score (which we term as an \textit{interaction function}). 

To estimate parameters $\Theta$, existing approaches generally follow the machine learning paradigm that optimizes an objective function. Two types of objective functions are most commonly used in literature --- pointwise loss~\cite{fastMF,Hu:2008} and pairwise loss~\cite{BPR,NTN}. As a natural extension of abundant work on explicit feedback~\cite{SVD++,DiscreteCF},
methods on pointwise learning usually follow a regression framework by minimizing the squared loss between $\hat{y}_{ui}$ and its target value $y_{ui}$. To handle the absence of negative data, they have either treated all unobserved entries as negative feedback, 
or sampled negative instances from unobserved entries~\cite{fastMF}. For pairwise learning~\cite{BPR,Wu:WSDM16}, the idea is that observed entries should be ranked higher than the unobserved ones. As such, instead of minimizing the loss between $\hat{y}_{ui}$ and $y_{ui}$, pairwise learning maximizes the margin between observed entry $\hat{y}_{ui}$ and unobserved entry $\hat{y}_{uj}$. 

Moving one step forward, our NCF framework parameterizes the interaction function $f$ using neural networks to estimate $\hat{y}_{ui}$. As such, it naturally supports both pointwise and pairwise learning. 

\subsection{Matrix Factorization}

\begin{figure}[t]
	\centering
	\begin{subfigure}[b]{0.2\textwidth}
		\centering
		\includegraphics[width=\textwidth]{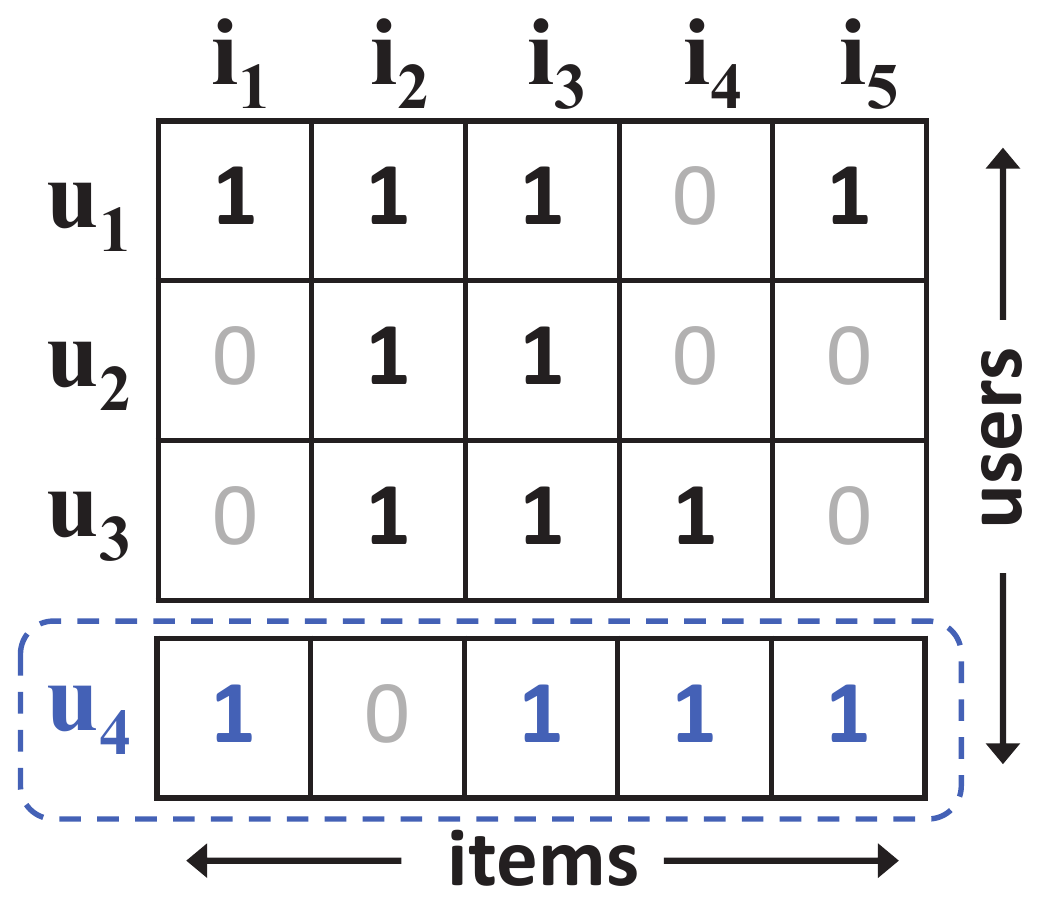}
		\vspace{-15pt}
		\caption{user--item matrix}
		\label{fig:user_item_matrix}
	\end{subfigure}  \hspace{+15pt}
	\begin{subfigure}[b]{0.18\textwidth}
		\centering
		\includegraphics[width=\textwidth]{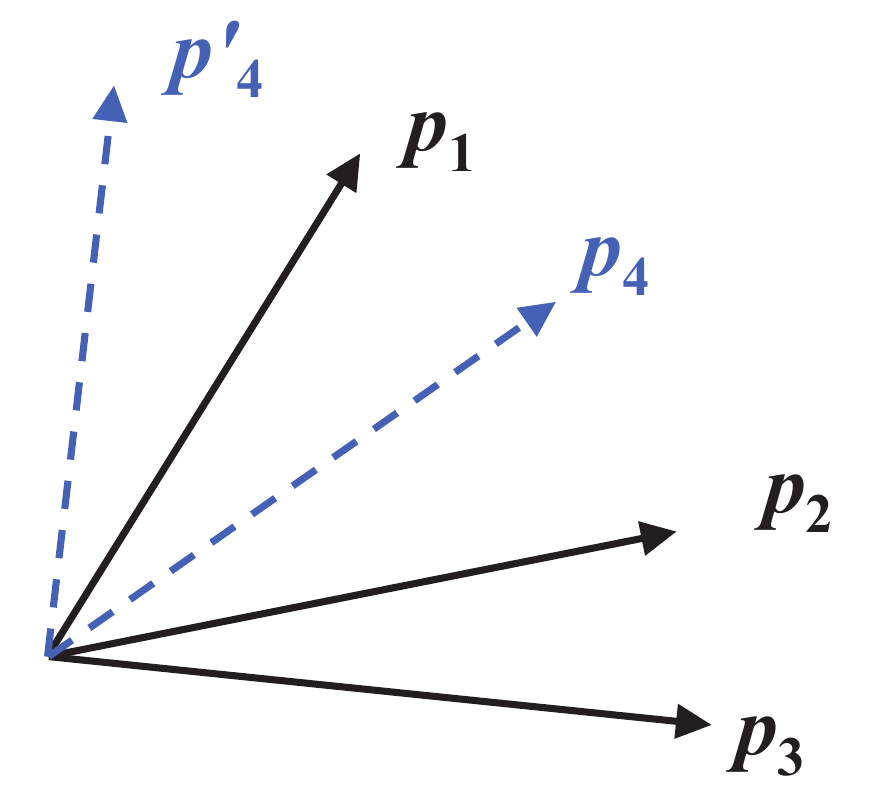}
		\vspace{-15pt}
		\caption{user latent space}
		\label{fig:user_latent_space}
	\end{subfigure} 
	\vspace{-5pt}
	\caption{An example illustrates MF's limitation. From data matrix (a), $u_4$ is most similar to $u_1$, followed by $u_3$, and lastly $u_2$.  However in the latent space (b), placing $\textbf{p}_4$ closest to $\textbf{p}_1$ makes $\textbf{p}_4$ closer to $\textbf{p}_2$ than $\textbf{p}_3$, incurring a large ranking loss.}
	\vspace{-10pt}
	\label{fig:mf_limitation}
\end{figure}

MF associates each user and item with a real-valued vector of latent features. Let $\textbf{p}_u$ and $\textbf{q}_i$  denote the latent vector for user $u$ and item $i$, respectively; MF estimates an interaction $y_{ui}$ as the inner product of $\textbf{p}_u$ and $\textbf{q}_i$: \vspace{-5pt}
\begin{equation}
\label{eq:MF}
	\hat{y}_{ui} = f(u,i|\textbf{p}_u,\textbf{q}_i) = \textbf{p}_u^T \textbf{q}_i = \sum_{k=1}^K p_{uk}  q_{ik},
\end{equation}
where $K$ denotes the dimension of the latent space. 
As we can see, MF models the two-way interaction of user and item latent factors, assuming each dimension of the latent space is independent of each other and linearly combining them with the same weight. 
As such, MF can be deemed as a linear model of latent factors. 

Figure~\ref{fig:mf_limitation} illustrates how the inner product function can limit the expressiveness of MF. 
There are two settings to be stated clearly beforehand to understand the example well. 
First, since MF maps users and items to the same latent space, the similarity between two users can also be measured with an inner product, or equivalently\footnote{Assuming latent vectors are of a unit length.}, the cosine of the angle between their latent vectors. 
Second, without loss of generality, we use the Jaccard coefficient\footnote{Let $\mathcal{R}_u$ be the set of items that user $u$ has interacted with, then the Jaccard similarity of users $i$ and $j$ is defined as $s_{ij} = \frac{|\mathcal{R}_i|\cap |\mathcal{R}_j|}{|\mathcal{R}_i|\cup |\mathcal{R}_j|}$.} as the ground-truth similarity of two users that MF needs to recover. 

Let us first focus on the first three rows (users) in Figure~\ref{fig:user_item_matrix}. It is easy to have $s_{23} (0.66) > s_{12} (0.5) > s_{13} (0.4)$. As such, the geometric relations of $\textbf{p}_1,\textbf{p}_2,$ and $\textbf{p}_3$ in the latent space can be plotted as in Figure~\ref{fig:user_latent_space}. Now, let us consider a new user $u_4$, whose input is given as the dashed line in Figure~\ref{fig:user_item_matrix}. We can have $s_{41} (0.6) > s_{43} (0.4) > s_{42} (0.2)$, meaning that $u_4$ is most similar to $u_1$, followed by $u_3$, and lastly $u_2$. However, if a MF model places $\textbf{p}_4$ closest to $\textbf{p}_1$ (the two options are shown in Figure~\ref{fig:user_latent_space} with dashed lines), it will result in $\textbf{p}_4$ closer to $\textbf{p}_2$ than $\textbf{p}_3$, which unfortunately will incur a large ranking loss. 

The above example shows the possible limitation of MF caused by the use of a simple and fixed inner product to estimate complex user--item interactions in the low-dimensional latent space. We note that one way to resolve the issue is to use a large number of latent factors $K$. However, it may adversely hurt the generalization of the model (\eg overfitting the data), especially in sparse settings~\cite{FM}. In this work, we address the limitation by learning the interaction function using DNNs from data. 

%% file: 3_method.tex
\section{Neural Collaborative Filtering}
\label{sec:method}
We first present the general NCF framework, elaborating how to learn NCF with a probabilistic model that emphasizes the binary property of implicit data. 
We then show that MF can be expressed and generalized under NCF. To explore DNNs for collaborative filtering, we then propose an instantiation of NCF, using a multi-layer perceptron (MLP) to learn the user--item interaction function. Lastly, we present a new neural matrix factorization model, which ensembles MF and MLP under the NCF framework; it unifies the strengths of linearity of MF and non-linearity of MLP for modelling the user--item latent structures. 

\begin{figure}[t]
	\centering
	\includegraphics[width=0.46\textwidth]{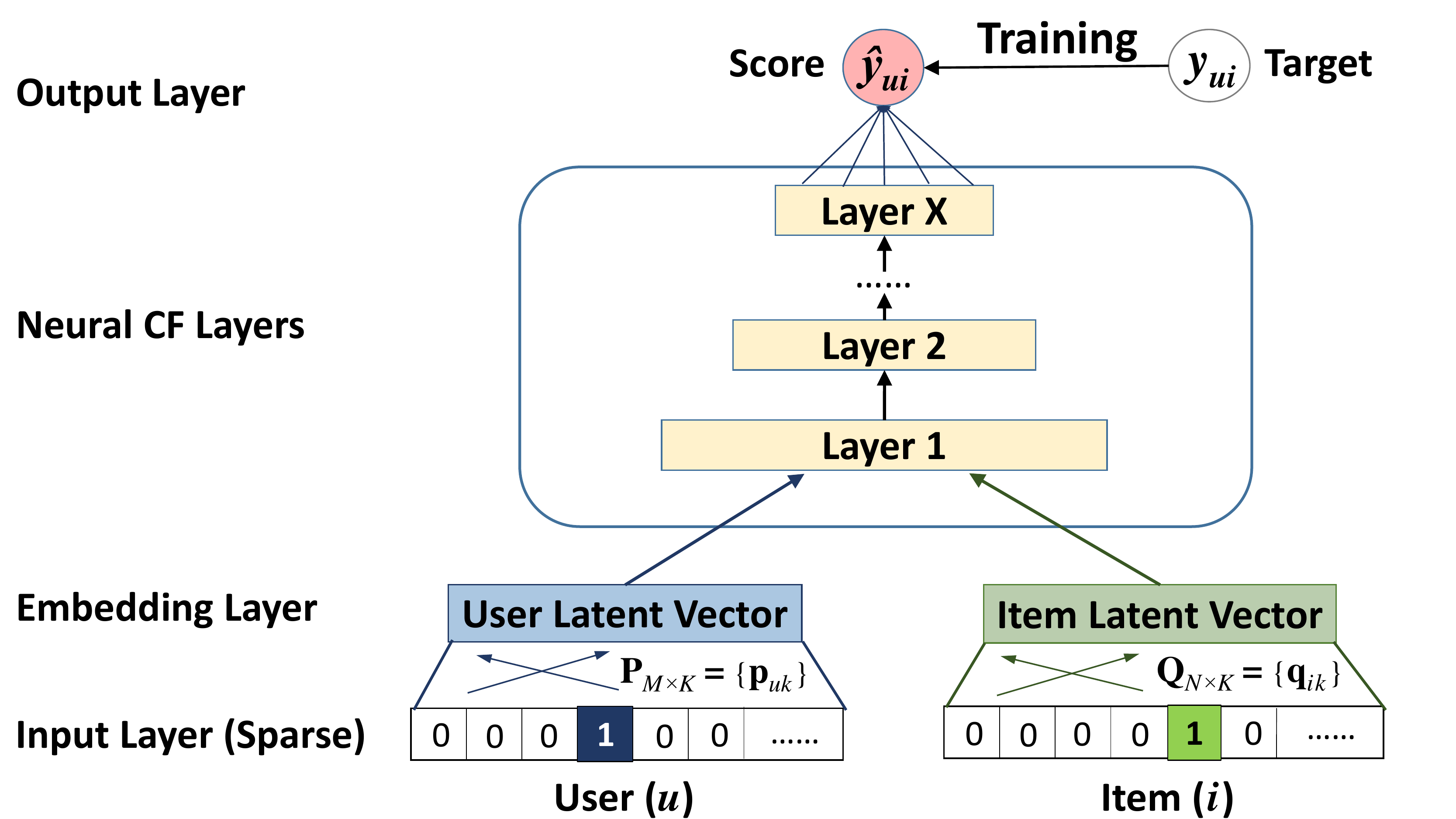}
	\vspace{-5pt}
	\caption{Neural collaborative filtering framework}
	\vspace{-10pt}
	\label{fig:framework}
\end{figure}

\subsection{General Framework}
To permit a full neural treatment of collaborative filtering, we adopt a multi-layer representation to model a user--item interaction $y_{ui}$ as shown in Figure~\ref{fig:framework}, where the output of one layer serves as the input of the next one. 
The bottom input layer consists of two feature vectors $\textbf{v}^U_u$ and $\textbf{v}^I_i$ that describe user $u$ and item $i$, respectively; they can be customized to support a wide range of modelling of users and items, such as context-aware~\cite{fastFM,iCD}, content-based~\cite{Chen:2016}, and neighbor-based~\cite{FM}. 
Since this work focuses on the pure collaborative filtering setting, we use only the identity of a user and an item as the input feature, transforming it to a binarized sparse vector with one-hot encoding. Note that with such a generic feature representation for inputs, 
our method can be easily adjusted to address the cold-start problem by using content features to represent users and items. 

Above the input layer is the embedding layer; it is a fully connected layer that projects the sparse representation to a dense vector. The obtained user (item) embedding can be seen as the latent vector for user (item) in the context of latent factor model.
The user embedding and item embedding are then fed into a multi-layer neural architecture, which we term as \textit{neural collaborative filtering layers},  to map the latent vectors to prediction scores. Each layer of the neural CF layers can be customized to discover certain latent structures of user--item interactions. 
The dimension of the last hidden layer $X$ determines the model's capability.
The final output layer is the predicted score $\hat{y}_{ui}$, and training is performed by minimizing the pointwise loss between $\hat{y}_{ui}$ and its target value $y_{ui}$. We note that another way to train the model is by performing pairwise learning, such as using the Bayesian Personalized Ranking~\cite{BPR} and margin-based loss~\cite{NTN}. As the focus of the paper is on the neural network modelling part, we leave the extension to pairwise learning of NCF as a future work. 

We now formulate the NCF's predictive model as
\begin{equation}
	\hat{y}_{ui} = f(\textbf{P}^T\textbf{v}^U_u, \textbf{Q}^T\textbf{v}^I_i | \textbf{P},\textbf{Q},\Theta_f),
\end{equation}
where $\textbf{P}\in\mathbb{R}^{M\times K}$ and $\textbf{Q}\in\mathbb{R}^{N\times K}$, denoting the latent factor matrix for users and items, respectively; and $\Theta_f$ denotes the model parameters of the interaction function $f$. Since the function $f$ is defined as a multi-layer neural network, it can be formulated as
\begin{equation}
	f(\textbf{P}^T\textbf{v}^U_u, \textbf{Q}^T\textbf{v}^I_i) = \phi_{out}(\phi_X( ... \phi_2 (\phi_1 (\textbf{P}^T\textbf{v}^U_u, \textbf{Q}^T\textbf{v}^I_i)) ... )),
\end{equation}
where $\phi_{out}$ and $\phi_x$ respectively denote the mapping function for the output layer and $x$-th neural collaborative filtering (CF) layer, and there are $X$ neural CF layers in total. 

\subsubsection{Learning NCF}
\label{ss:learning_ncf}
To learn model parameters, existing pointwise methods~\cite{fastMF,Wang_anchor} largely perform a regression with squared loss:
\begin{equation}
\label{eq:squared_loss}
L_{sqr} = \sum_{(u,i)\in\mathcal{Y}\cup\mathcal{Y}^-} w_{ui} (y_{ui} - \hat{y}_{ui})^2,
\end{equation}
where $\mathcal{Y}$ denotes the set of observed interactions in $\textbf{Y}$, and  $\mathcal{Y}^-$ denotes the set of negative instances, which can be all (or sampled from) unobserved interactions; and $w_{ui}$ is a hyper-parameter denoting the weight of training instance $(u,i)$. While the squared loss can be explained by assuming that observations are generated from a Gaussian distribution~\cite{PMF}, we point out that it may not tally well with implicit data.
This is because for implicit data, the target value $y_{ui}$ is a binarized 1 or 0 denoting whether $u$ has interacted with $i$. In what follows, we present a probabilistic approach for learning the pointwise NCF that pays special attention to the binary property of implicit data. 

Considering the one-class nature of implicit feedback, we can view the value of $y_{ui}$ as a label --- 1 means item $i$ is relevant to $u$, and 0 otherwise. The prediction score $\hat{y}_{ui}$ then represents how likely $i$ is relevant to $u$. To endow NCF with such a probabilistic explanation, we need to constrain the output $\hat{y}_{ui}$ in the range of $[0,1]$, which can be easily achieved by using a probabilistic function (\eg the \textit{Logistic} or \textit{Probit} function) as the activation function for the output layer $\phi_{out}$. With the above settings, we then define the likelihood function as
\begin{equation}
\label{eq:likelihood}
p(\mathcal{Y},\mathcal{Y}^- | \textbf{P},\textbf{Q},\Theta_f) = \prod_{(u,i)\in\mathcal{Y}} \hat{y}_{ui}  \prod_{(u,j)\in\mathcal{Y}^-} (1 - \hat{y}_{uj}).
\end{equation}
Taking the negative logarithm of the likelihood, we reach
\begin{equation}
\label{eq:objective}
\begin{aligned}
L &= -\sum_{(u,i)\in\mathcal{Y}}\log \hat{y}_{ui} - \sum_{(u,j)\in\mathcal{Y}^-}\log (1 - \hat{y}_{uj}) \\
&= -\sum_{(u,i)\in\mathcal{Y}\cup\mathcal{Y}^-} y_{ui}\log \hat{y}_{ui} + (1 - y_{ui}) \log (1 - \hat{y}_{ui}).
\end{aligned}
\end{equation}
This is the objective function to minimize for the NCF methods, and its optimization can be done by performing stochastic gradient descent~(SGD). 
Careful readers might have realized that it is the same as the \textit{binary cross-entropy loss}, also known as \textit{log loss}. By employing a probabilistic treatment for NCF, we address recommendation with implicit feedback as a binary classification problem. As the classification-aware log loss has rarely been investigated in recommendation literature, we explore it in this work and empirically show its effectiveness in Section~\ref{ss:logloss}. For the negative instances $\mathcal{Y}^-$, we uniformly sample them from unobserved interactions in each iteration and control the sampling ratio \wrt the number of observed interactions. While a non-uniform sampling strategy (\eg item popularity-biased~\cite{fastMF,HeSIGIR2014}) might further improve the performance, we leave the exploration as a future work.



\subsection{Generalized Matrix Factorization (GMF)}
We now show how MF can be interpreted as a special case of our NCF framework. As MF is the most popular model for recommendation and has been investigated extensively in literature, being able to recover it allows NCF to mimic a large family of factorization models~\cite{FM}. 

Due to the one-hot encoding of user (item) ID of the input layer, the obtained embedding vector can be seen as the latent vector of user (item). Let the user latent vector $\textbf{p}_u$ be $\textbf{P}^T\textbf{v}^U_u$ and item latent vector $\textbf{q}_i$ be $\textbf{Q}^T\textbf{v}^I_i$. We define the mapping function of the first neural CF layer as
\begin{equation}
\phi_1(\textbf{p}_u, \textbf{q}_i) = \textbf{p}_u\odot
 \textbf{q}_i,
\end{equation}
where $\odot$ denotes the element-wise product of vectors. We then project the vector to the output layer:
\begin{equation}
\hat{y}_{ui} =  a_{out}(\textbf{h}^T(\textbf{p}_u\odot
\textbf{q}_i)),
\end{equation}
where $a_{out}$ and $\textbf{h}$ denote the activation function and edge weights of the output layer, respectively. Intuitively, if we use an identity function for $a_{out}$ and enforce $\textbf{h}$ to be a uniform vector of 1, we can exactly recover the MF model.

Under the NCF framework, MF can be easily generalized and extended. For example, if we allow $\textbf{h}$ to be learnt from data without the uniform constraint, it will result in a variant of MF that allows varying importance of latent dimensions. And if we use a non-linear function for $a_{out}$, it will generalize MF to a non-linear setting which might be more expressive than the linear MF model. In this work, we implement a generalized version of MF under NCF that uses the sigmoid function $\sigma(x)=1/(1+e^{-x})$ as $a_{out}$ and learns $\textbf{h}$ from data with the log loss~(Section~\ref{ss:learning_ncf}). We term it as GMF, short for \textit{Generalized Matrix Factorization}.

\subsection{Multi-Layer Perceptron (MLP)}
\label{ss:mlp}
Since NCF adopts two pathways to model users and items, it is intuitive to combine the features of two pathways by concatenating them. This design has been widely adopted in multimodal deep learning work~\cite{zhang2014start,srivastava2012multimodal}. However, simply a vector concatenation does not account for any interactions between user and item latent features, which is insufficient for modelling the collaborative filtering effect. To address this issue, we propose to add hidden layers on the concatenated vector, using a standard MLP to learn the interaction between user and item latent features. 
In this sense, we can endow the model a large level of flexibility and non-linearity to learn the interactions between $\textbf{p}_u$ and $\textbf{q}_i$, rather than the way of GMF that uses only a fixed element-wise product on them. 
More precisely, the MLP model under our NCF framework is defined as
\begin{equation}
\begin{aligned}
	\textbf{z}_1 &= \phi_1(\textbf{p}_u, \textbf{q}_i) = 
	\begin{bmatrix} \textbf{p}_u \\ \textbf{q}_i \end{bmatrix} , \\
	\phi_2 (\textbf{z}_1) &= a_2(\textbf{W}_2^T \textbf{z}_1 + \textbf{b}_2), \\
	&...... \\
	\phi_L(\textbf{z}_{L-1}) &= a_L(\textbf{W}_L^T \textbf{z}_{L-1} + \textbf{b}_L), \\
	\hat{y}_{ui} &= \sigma(\textbf{h}^T \phi_L(\textbf{z}_{L-1})),
\end{aligned}
\end{equation}
where $\textbf{W}_x$, $\textbf{b}_x$, and $a_{x}$ denote the weight matrix, bias vector, and activation function for the $x$-th layer's perceptron, respectively.
For activation functions of MLP layers, one can freely choose sigmoid, hyperbolic tangent~(tanh), and Rectifier~(ReLU), among others. 
We would like to analyze each function: 
1) The sigmoid function restricts each neuron to be in (0,1), which may limit the model's performance; and it is known to suffer from saturation, where neurons stop learning when their output is near either 0 or 1. 2) Even though tanh is a better choice and has been widely adopted~\cite{Elkahky:2015:MVDL,Wu:WSDM16}, it only alleviates the issues of sigmoid to a certain extent, since it can be seen as a rescaled version of sigmoid~(tanh$(x/2)=2\sigma(x)-1$). And 3) as such, we opt for ReLU, which is more biologically
plausible and proven to be non-saturated~\cite{glorot2011deep};
moreover, it encourages sparse activations, being well-suited for sparse data and making the model less likely to be overfitting. Our empirical results
show that ReLU yields slightly better performance than tanh, which in turn is significantly better than sigmoid. 

As for the design of network structure, a common solution is to follow a tower pattern, where the bottom layer is the widest and each successive layer has a smaller number of neurons (as in Figure~\ref{fig:framework}). The premise is that by using a small number of hidden units for higher layers, they can learn more abstractive features of data~\cite{cvpr16best}. We empirically implement the tower structure, halving the layer size for each successive higher layer. 

\subsection{Fusion of GMF and MLP}
\label{ss:NeuMF}
So far we have developed two instantiations of NCF --- GMF that applies a linear kernel to model the latent feature interactions, and MLP that uses a non-linear kernel to learn the interaction function from data. 
The question then arises: how can we fuse GMF and MLP under the NCF framework, so that they can mutually reinforce each other to better model the complex user-iterm interactions? 

\begin{figure}[t]
	\centering
	\includegraphics[width=0.50\textwidth]{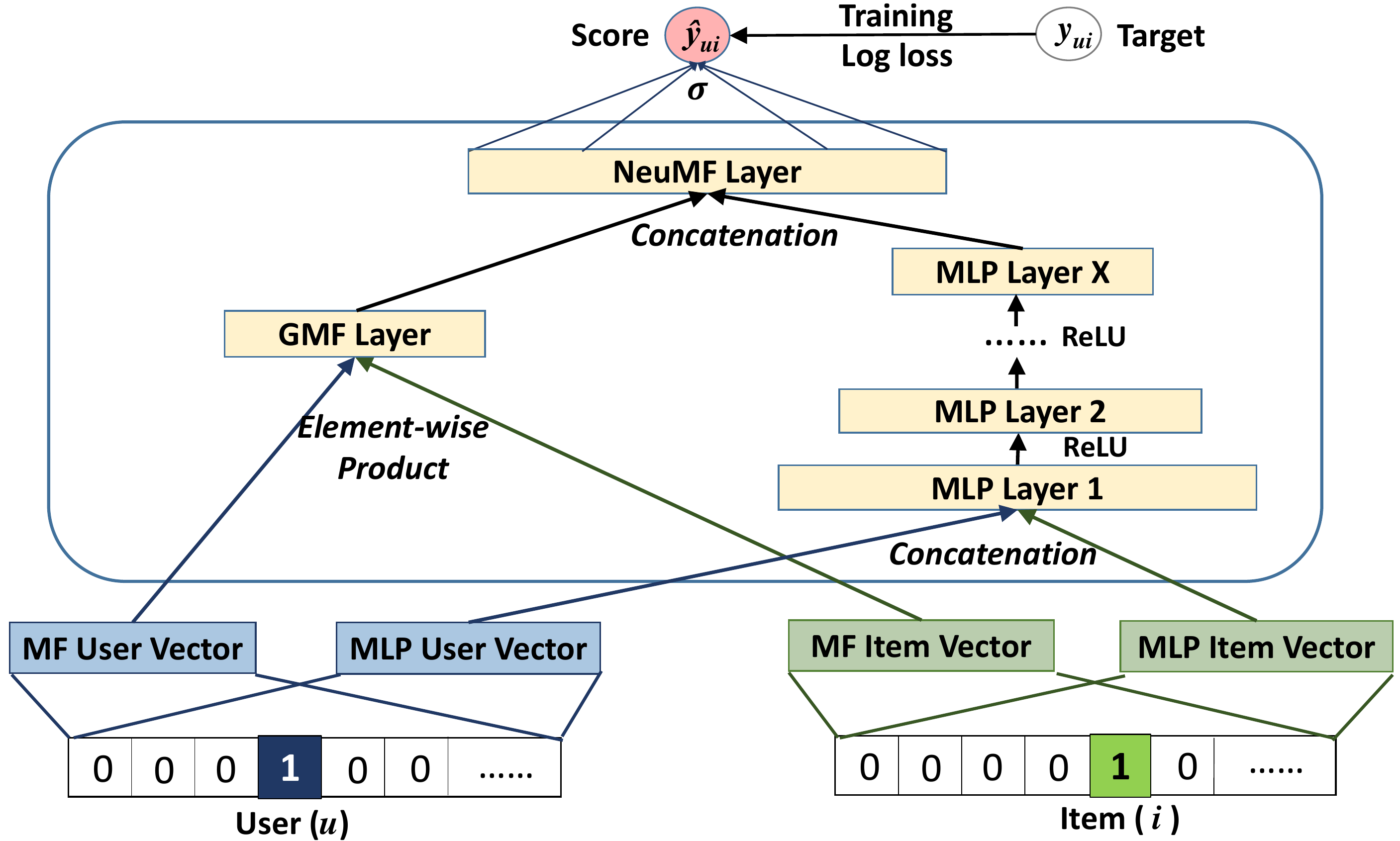}
	\vspace{-15pt}
	\caption{Neural matrix factorization model}
	\vspace{-15pt}
	\label{fig:NeuMF}
\end{figure}

A straightforward solution is to let GMF and MLP share the same embedding layer, and then combine the outputs of their interaction functions. This way shares a similar spirit with the well-known Neural Tensor Network~(NTN)~\cite{NTN}. Specifically, the model for combining GMF with a one-layer MLP can be formulated as
\begin{equation}
	\hat{y}_{ui} = \sigma(\textbf{h}^T a(\textbf{p}_u \odot \textbf{q}_i + \textbf{W}\begin{bmatrix} \textbf{p}_u \\ \textbf{q}_i \end{bmatrix} + \textbf{b})).
\end{equation}

However, sharing embeddings of GMF and MLP might limit the performance of the fused model. 
For example, it implies that GMF and MLP must use the same size of embeddings; for datasets where the optimal embedding size of the two models varies a lot, this solution may fail to obtain the optimal ensemble. 


To provide more flexibility to the fused model, we allow GMF and MLP to learn separate embeddings, and combine the two models by concatenating their last hidden layer. Figure~\ref{fig:NeuMF} illustrates our proposal, the formulation of which is given as follows
\begin{equation}
\begin{aligned}
\small
	\phi^{GMF} &= \textbf{p}_u^{G} \odot \textbf{q}_i^{G}, \\
	\phi^{MLP} &= a_L(\textbf{W}_L^T (a_{L-1}( ...a_2(\textbf{W}_2^T \begin{bmatrix} \textbf{p}^M_u \\ \textbf{q}^M_i \end{bmatrix} + \textbf{b}_2) ...)) + \textbf{b}_L), \\
	\hat{y}_{ui} &= \sigma(\textbf{h}^T \begin{bmatrix}
	\phi^{GMF}\\ \phi^{MLP}
	\end{bmatrix} ),
\end{aligned}
\end{equation}
where $\textbf{p}_u^G$ and $\textbf{p}_u^M$ denote the user embedding for GMF and MLP parts, respectively; and similar notations of $\textbf{q}_i^G$ and $\textbf{q}_i^M$ for item embeddings. As discussed before, we use ReLU as the activation function of MLP layers. This model combines the linearity of MF and non-linearity of DNNs for modelling user--item latent structures. 
We dub this model ``NeuMF'', short for \textit{Neural Matrix Factorization}. The derivative of the model \wrt each model parameter can be calculated with standard back-propagation, which is omitted here due to space limitation. 

\subsubsection{Pre-training}
\label{ss:pre-training}
Due to the non-convexity of the objective function of NeuMF, gradient-based optimization methods only find locally-optimal solutions. 
It is reported that the initialization plays an important role for the convergence and performance of deep learning models~\cite{Erhan:2010}.
Since NeuMF is an ensemble of GMF and MLP, we propose to initialize NeuMF using the pre-trained models of GMF and MLP. 

We first train GMF and MLP with random initializations until convergence. We then use their model parameters as the initialization for the corresponding parts of NeuMF's parameters. The only tweak is on the output layer, where we concatenate weights of the two models with
\begin{equation}
	 \textbf{h} \leftarrow \begin{bmatrix}
	 \alpha\textbf{h}^{GMF} \\ (1-\alpha)\textbf{h}^{MLP}
	 \end{bmatrix},
\end{equation}
where $\textbf{h}^{GMF}$ and $\textbf{h}^{MLP}$ denote the $\textbf{h}$ vector of the pre-trained GMF and MLP model, respectively; and $\alpha$ is a hyper-parameter determining the trade-off between the two pre-trained models. 

For training GMF and MLP from scratch, we adopt the Adaptive Moment Estimation (Adam)~\cite{adam}, which adapts the learning rate for each parameter 
by performing smaller updates for frequent and larger updates for infrequent parameters. The Adam method yields faster convergence for both models than the vanilla SGD and relieves the pain of tuning the learning rate. After feeding pre-trained parameters into NeuMF, we optimize it with the vanilla SGD, rather than Adam. This is because Adam needs to save momentum information for updating parameters properly. As we initialize NeuMF with pre-trained model parameters only and forgo saving the momentum information, it is unsuitable to further optimize NeuMF with momentum-based methods. 

%% file: 4_experiment.tex
\section{Experiments}
\label{sec:experiments}

In this section, we conduct experiments with the aim of answering the following research questions: 
\begin{itemize}
	\item[\textbf{RQ1}] Do our proposed NCF methods outperform the state-of-the-art implicit collaborative filtering methods? 
	\item[\textbf{RQ2}] How does our proposed optimization framework (log loss with negative sampling) work for the recommendation task?
	\item[\textbf{RQ3}] Are deeper layers of hidden units helpful for learning from user--item interaction data?
\end{itemize}
In what follows, we first present the experimental settings, followed by answering the above three research questions. 


\subsection{Experimental Settings}
\noindent\textbf{Datasets.} We experimented with two publicly accessible datasets: MovieLens\footnote{\url{http://grouplens.org/datasets/movielens/1m/}} and Pinterest\footnote{\url{https://sites.google.com/site/xueatalphabeta/academic-projects}}. The characteristics of the two datasets are summarized in Table \ref{tab:dataset}. 

\textbf{1. MovieLens}. This movie rating dataset has been widely used to evaluate collaborative filtering algorithms. We used the version containing one million ratings, where each user has at least 20 ratings. 
While it is an explicit feedback data, we have intentionally chosen it to investigate the performance of learning from the implicit signal~\cite{SVD++} of explicit feedback. To this end, we transformed it into implicit data, where each entry is marked as 0 or 1 indicating whether the user has rated the item. 

\textbf{2. Pinterest}. This implicit feedback data is constructed by \cite{Geng:2015} for evaluating content-based image recommendation. The original data is very large but highly sparse. For example, over $20\%$ of users have only one pin, making it difficult to evaluate collaborative filtering algorithms. As such, we filtered the dataset in the same way as the MovieLens data that retained only users with at least 20 interactions (pins). This results in a subset of the data that contains $55,187$ users and $1,500,809$ interactions. Each interaction denotes whether the user has pinned the image to her own board. \\
\vspace{-5pt}

\begin{table}[t]
	\begin{center}
		\caption{\textbf{Statistics of the evaluation datasets.}}
		\vspace{-8pt}
		\small
		\label{tab:dataset}
		\begin{tabular}{ | l | c | c | c | c | }
			\hline
			\textbf{Dataset} & \textbf{Interaction\#} & \textbf{Item\#} & \textbf{User\#}  & \textbf{Sparsity} \\ \hline
			MovieLens	& 1,000,209 & 3,706 & 6,040  & 95.53\% \\ \hline
			Pinterest	& 1,500,809	& 9,916 & 55,187 & 99.73\% \\ \hline
		\end{tabular}
		\vspace{-25pt}
	\end{center}
\end{table}

\noindent\textbf{Evaluation Protocols.} To evaluate the performance of item recommendation, we adopted the \textit{leave-one-out} evaluation, which has been widely used in literature~\cite{iCD,fastMF,BPR}. For each user, we held-out her latest interaction as the test set and utilized the remaining data for training. Since it is too time-consuming to rank all items for every user during evaluation, we followed the common strategy~\cite{Elkahky:2015:MVDL,SVD++} that randomly samples 100 items that are not interacted by the user, ranking the test item among the 100 items. The performance of a ranked list is judged by \textit{Hit Ratio} (HR) and \textit{Normalized Discounted Cumulative Gain} (NDCG)~\cite{TriRank}. Without special mention, we truncated the ranked list at $10$ for both metrics. As such, the HR intuitively measures whether the test item is present on the top-10 list, and the NDCG accounts for the position of the hit by assigning higher scores to hits at top ranks. We calculated both metrics for each test user and reported the average score.  \\
\vspace{-5pt}

\begin{figure*}[t]
	\centering
	\begin{subfigure}[b]{0.25\textwidth}
		\centering
		\includegraphics[width=\textwidth]{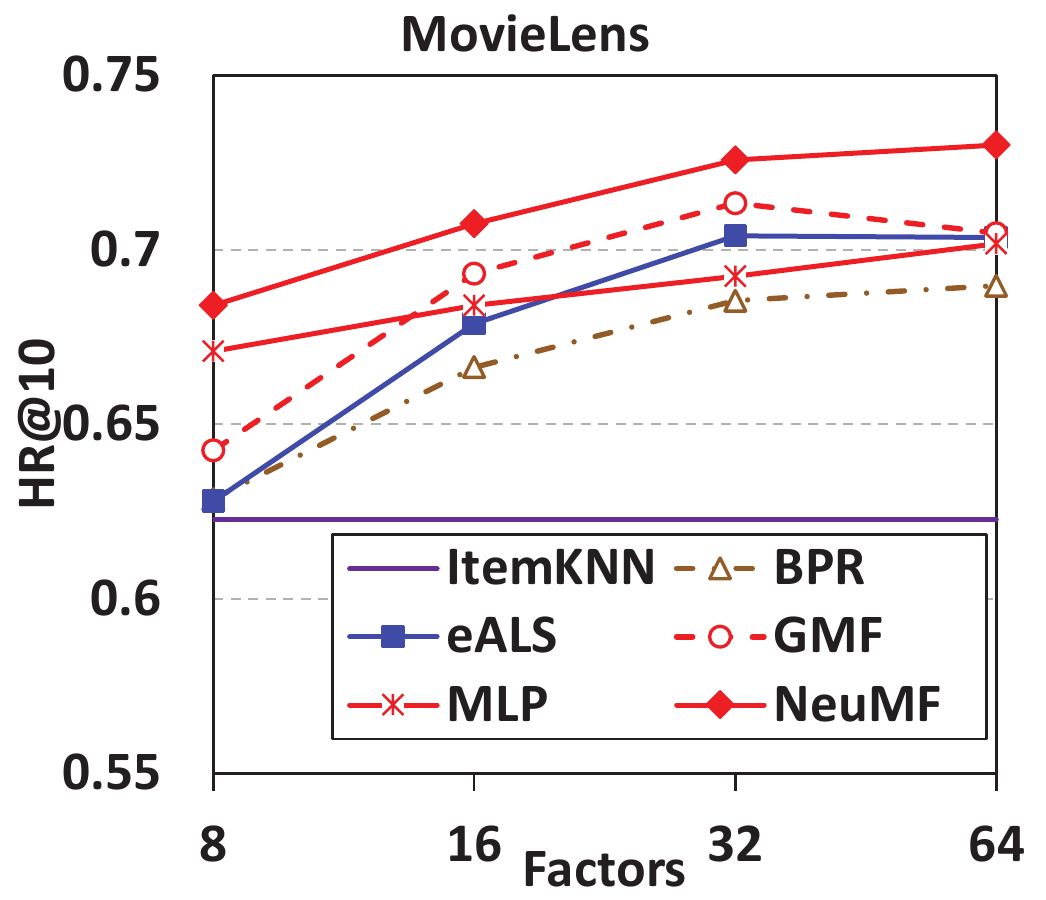}
		\vspace{-15pt}
		\caption{MovieLens --- HR@10}
		\label{fig:ml-hr-factors}
	\end{subfigure} \hspace{-7pt}
	\begin{subfigure}[b]{0.25\textwidth}
		\centering
		\includegraphics[width=\textwidth]{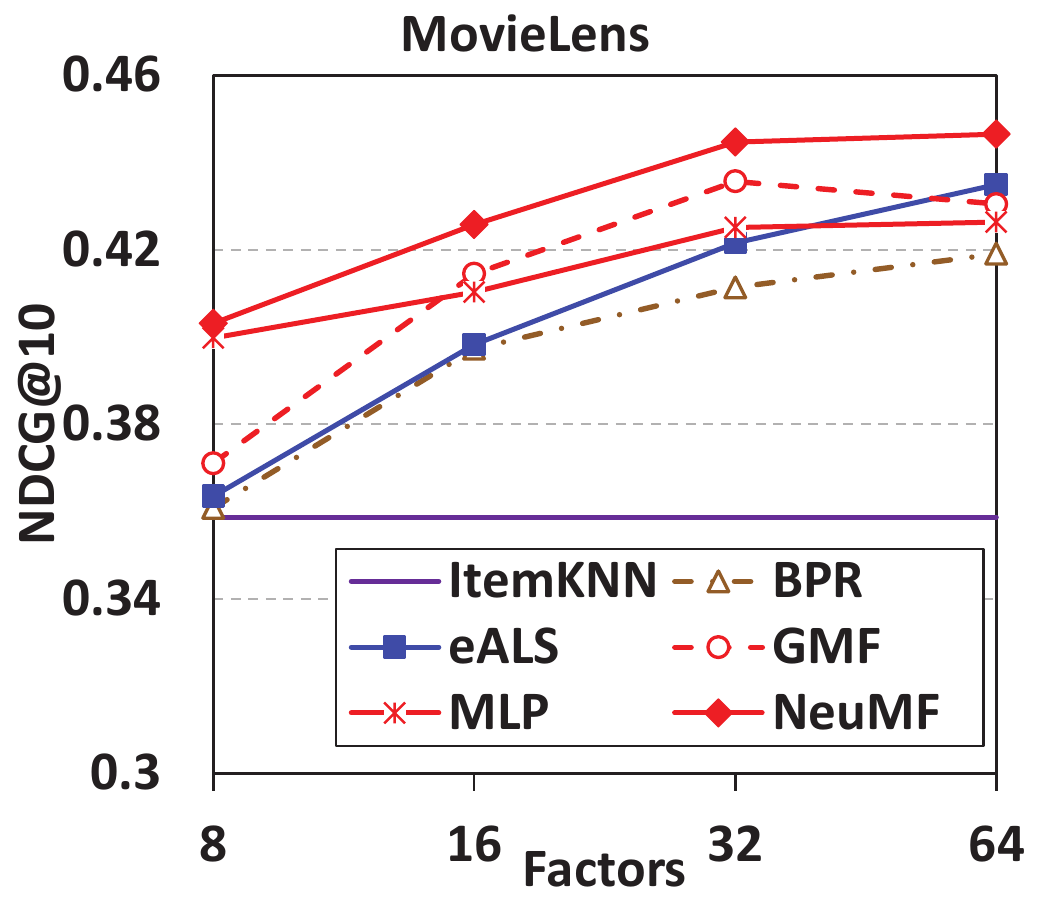}
		\vspace{-15pt}
		\caption{MovieLens --- NDCG@10}
		\label{fig:ml-ndcg-factors}
	\end{subfigure} \hspace{-7pt}
	\begin{subfigure}[b]{0.25\textwidth}
		\centering
		\includegraphics[width=\textwidth]{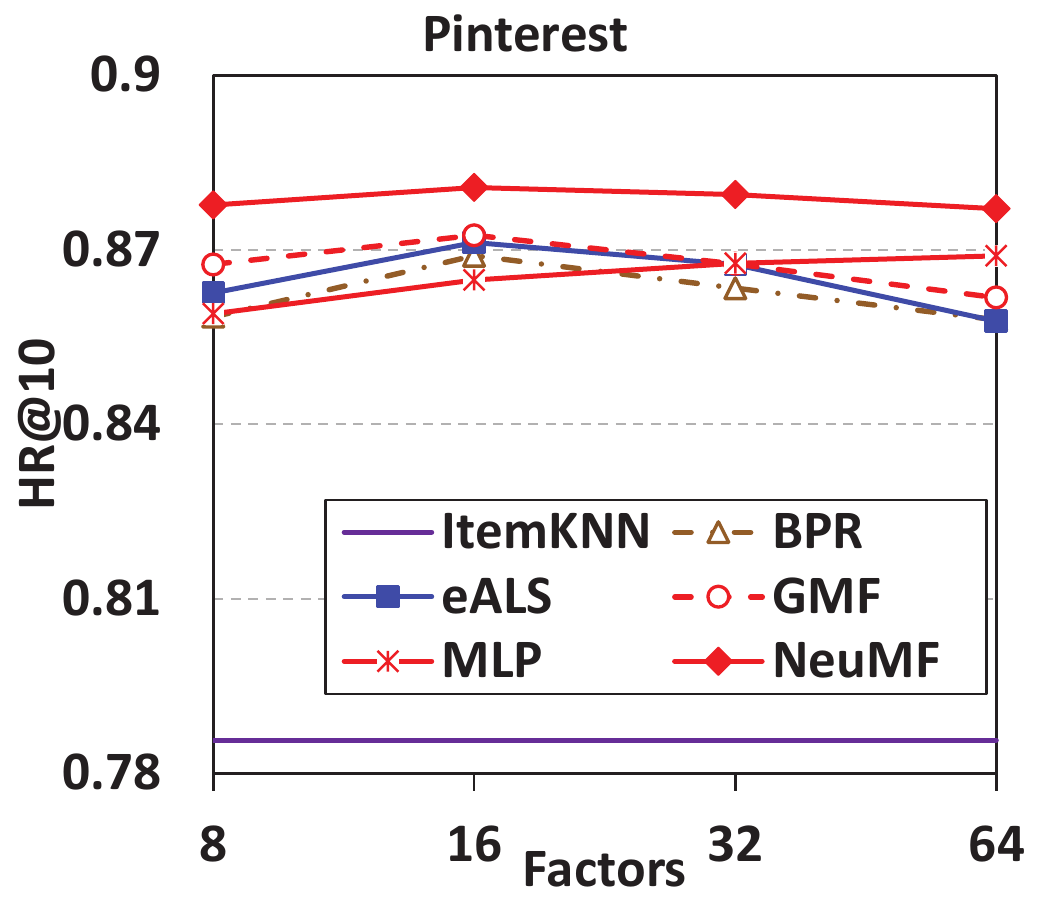}
		\vspace{-15pt}
		\caption{Pinterest --- HR@10}
		\label{fig:pinterest-hr-factors}
	\end{subfigure} \hspace{-7pt}
	\begin{subfigure}[b]{0.25\textwidth}
		\centering
		\includegraphics[width=\textwidth]{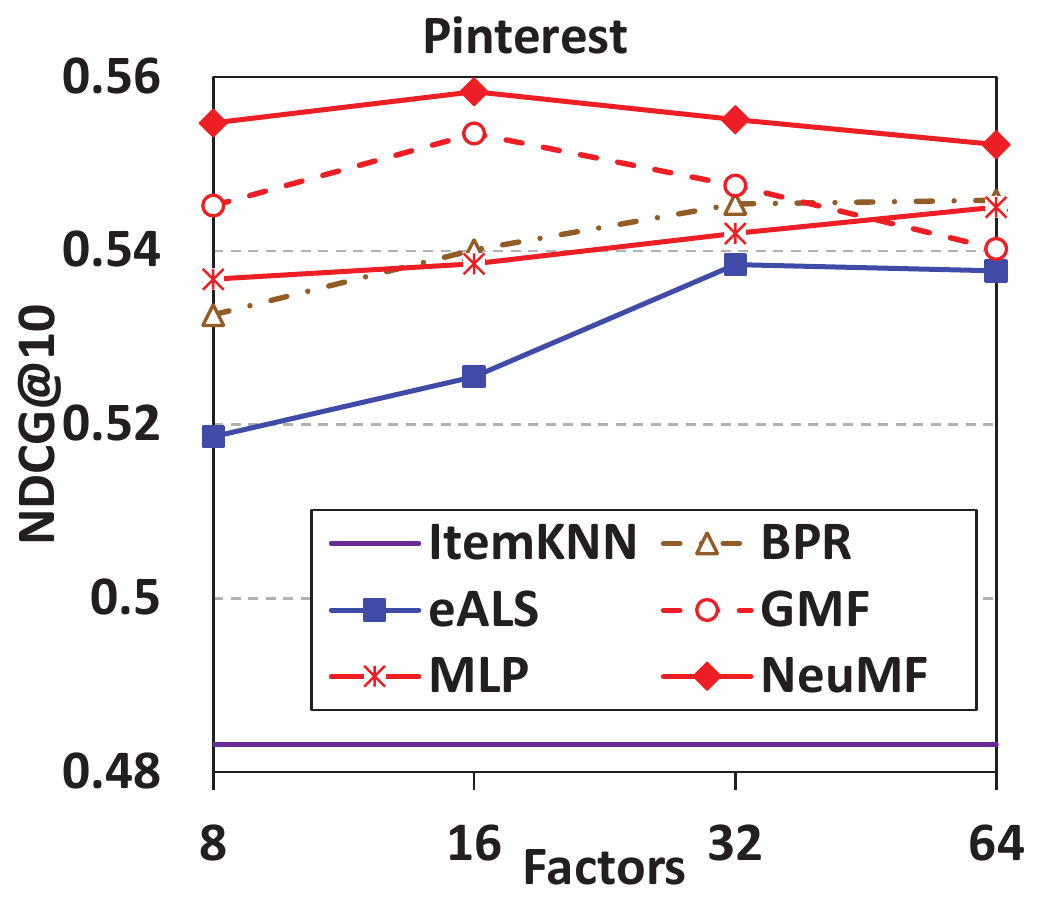}
		\vspace{-15pt}
		\caption{Pinterest --- NDCG@10}
		\label{fig:pinterest-ndcg-factors}
	\end{subfigure} \hspace{-7pt}
	\vspace{-5pt}
	\caption{Performance of HR@10 and NDCG@10 \wrt the number of predictive factors on the two datasets. }
	\label{fig:performance-factors}
\end{figure*}

\begin{figure*}[t]
	\centering
	\begin{subfigure}[b]{0.25\textwidth}
		\centering
		\includegraphics[width=\textwidth]{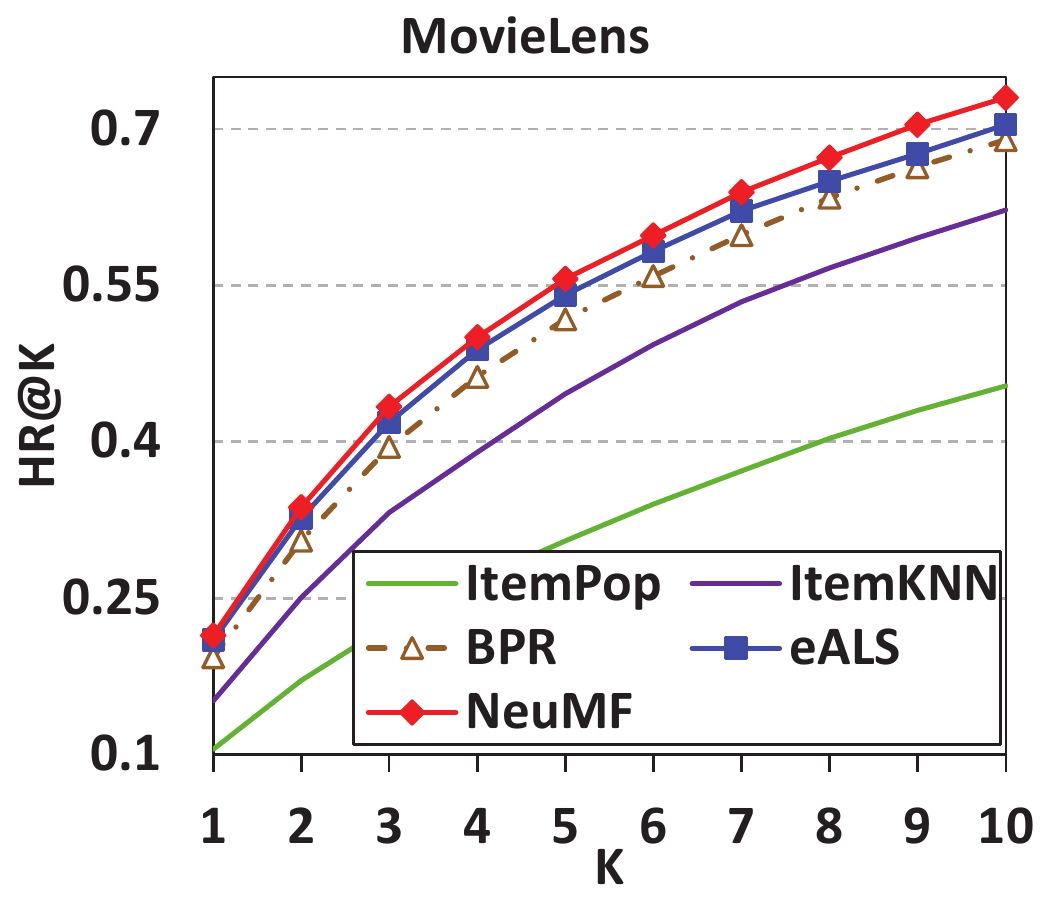}
		\vspace{-15pt}
		\caption{MovieLens --- HR@K}
		\label{fig:ml-hr-topk}
	\end{subfigure} \hspace{-7pt}
	\begin{subfigure}[b]{0.25\textwidth}
		\centering
		\includegraphics[width=\textwidth]{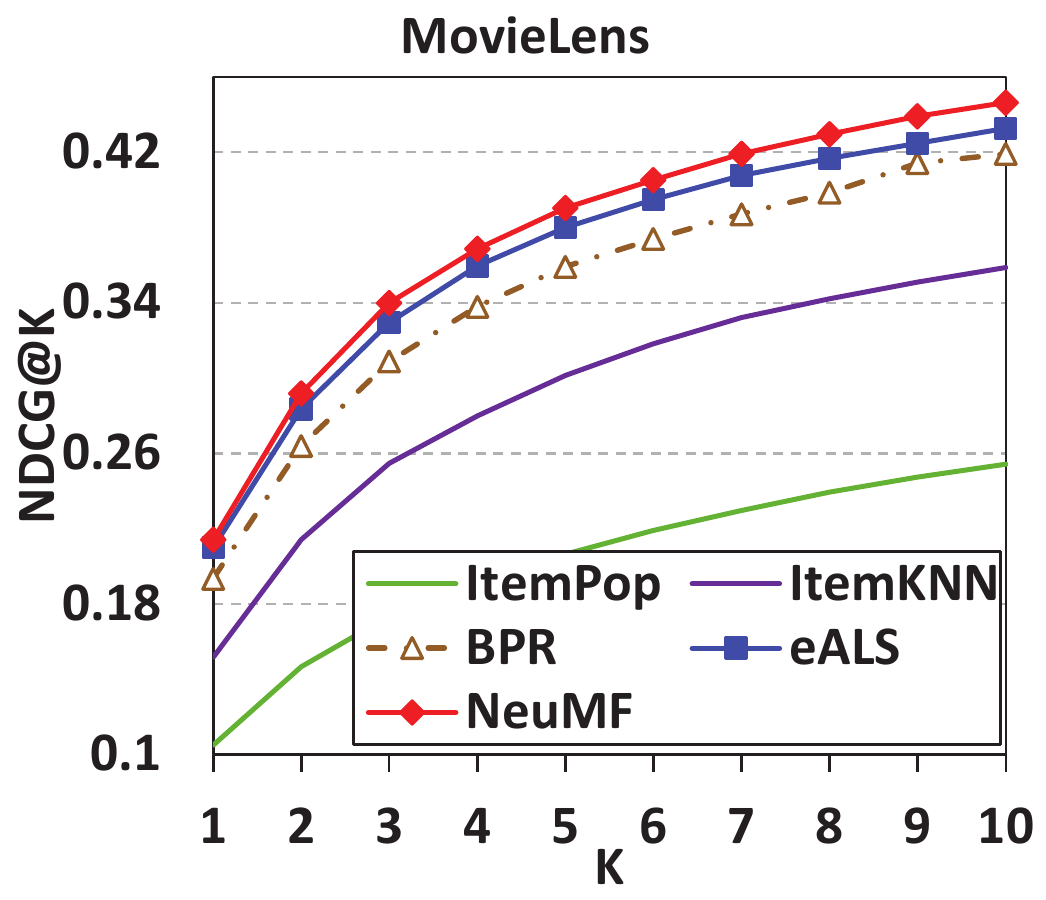}
		\vspace{-15pt}
		\caption{MovieLens --- NDCG@K}
		\label{fig:ml-ndcg-topk}
	\end{subfigure} \hspace{-7pt}
	\begin{subfigure}[b]{0.25\textwidth}
		\centering
		\includegraphics[width=\textwidth]{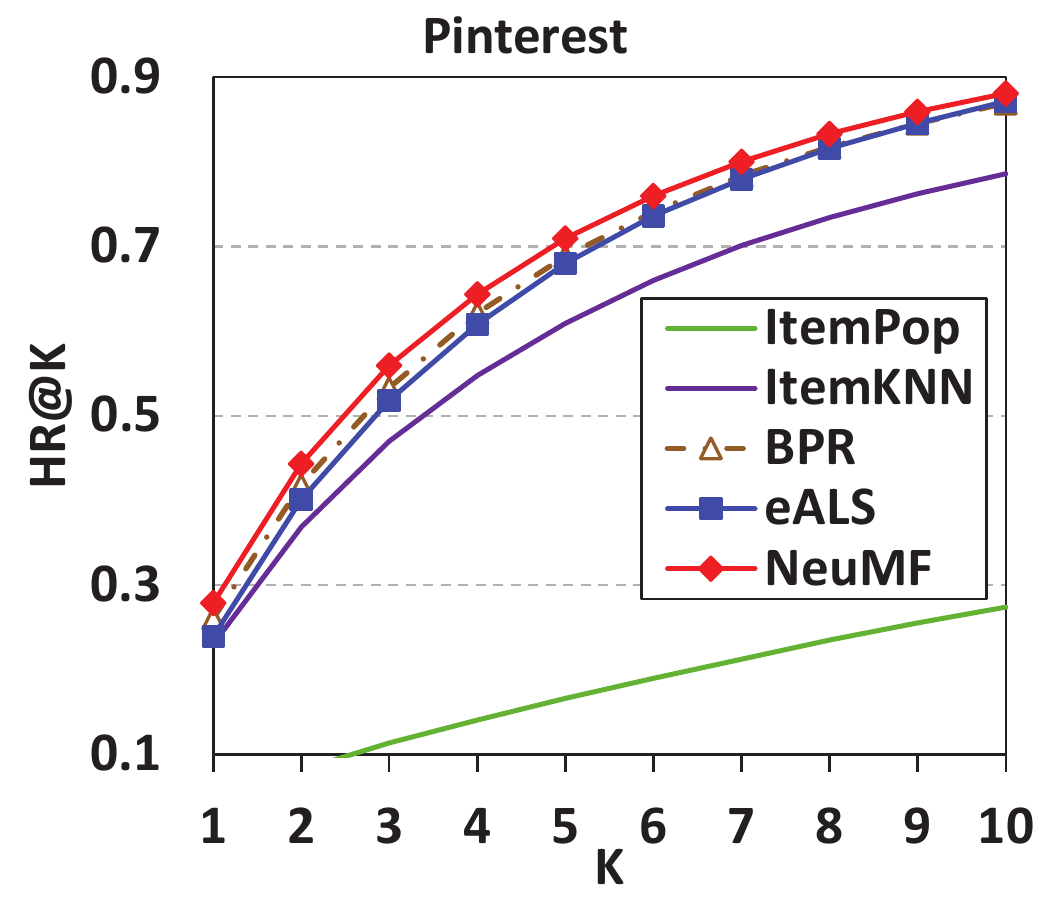}
		\vspace{-15pt}
		\caption{Pinterest --- HR@K}
		\label{fig:pinterest-hr-topk}
	\end{subfigure} \hspace{-7pt}
	\begin{subfigure}[b]{0.25\textwidth}
		\centering
		\includegraphics[width=\textwidth]{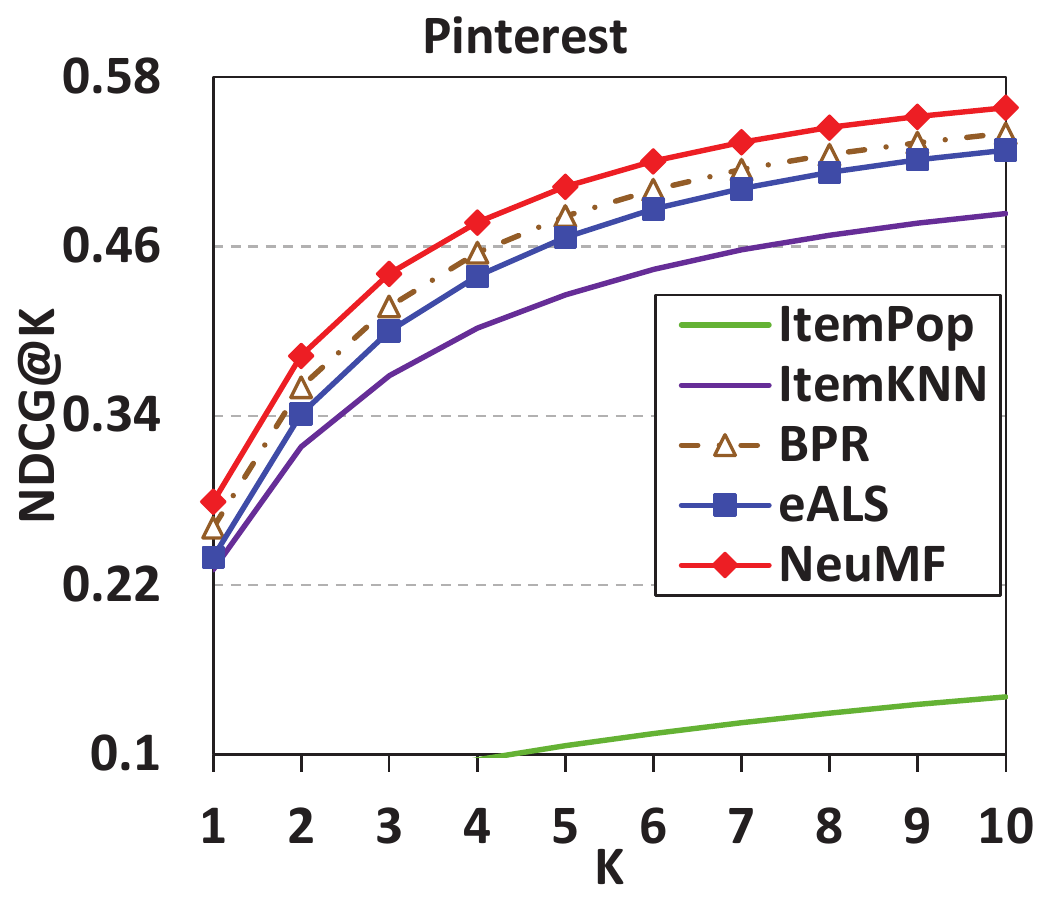}
		\vspace{-15pt}
		\caption{Pinterest --- NDCG@K}
		\label{fig:pinterest-ndcg-topk}
	\end{subfigure} \hspace{-7pt}
	\vspace{-5pt}
	\caption{Evaluation of Top-$K$ item recommendation where $K$ ranges from $1$ to $10$ on the two datasets.}
	\label{fig:performance-topk}
\end{figure*}

\noindent\textbf{Baselines.} 
We compared our proposed NCF methods (GMF, MLP and NeuMF) with the following methods: 

- \textbf{ItemPop}. Items are ranked by their popularity judged by the number of interactions. This is a non-personalized method to benchmark the recommendation performance~\cite{BPR}. 

- \textbf{ItemKNN}~\cite{ItemCF}. This is the standard item-based collaborative filtering method. 
We followed the setting of \cite{Hu:2008} to adapt it for implicit data. 

- \textbf{BPR}~\cite{BPR}. This method optimizes the MF model of Equation~\ref{eq:MF} with a pairwise ranking loss, which is tailored to learn from implicit feedback. It is a highly competitive baseline for item recommendation.  We used a fixed learning rate, varying it and reporting the best performance.

- \textbf{eALS}~\cite{fastMF}. This is a state-of-the-art MF method for item recommendation. It optimizes the squared loss of Equation~\ref{eq:squared_loss}, treating all unobserved interactions as negative instances and weighting them non-uniformly by the item popularity. 
Since eALS shows superior performance over the uniform-weighting method WMF~\cite{Hu:2008}, we do not further report WMF's performance. \vspace{+2pt}

As our proposed methods aim to model the relationship between users and items, we mainly compare with user--item models. We leave out the comparison with item--item models, such as SLIM~\cite{SLIM}
and CDAE~\cite{Wu:WSDM16}, because the performance difference may be caused by the user models for personalization (as they are item--item model). \\ \vspace{-5pt}

\noindent\textbf{Parameter Settings.} We implemented our proposed methods based on Keras\footnote{\url{https://github.com/hexiangnan/neural_collaborative_filtering}}. 
To determine hyper-parameters of NCF methods, we randomly sampled one interaction for each user as the validation data and tuned hyper-parameters on it. 
All NCF models are learnt by optimizing the log loss of Equation~\ref{eq:objective}, where we sampled four negative instances per positive instance. 
For NCF models that are trained from scratch, we randomly initialized model parameters with a Gaussian distribution (with a mean of $0$ and standard deviation of $0.01$), optimizing the model with mini-batch Adam~\cite{adam}.
We tested the batch size of $[128,256,512,1024]$, and the learning rate of [0.0001 ,0.0005, 0.001, 0.005]. 
Since the last hidden layer of NCF determines the model capability, we term it as \textit{predictive factors} and evaluated the factors of $[8,16,32,64]$. It is worth noting that large factors may cause overfitting and degrade the performance. 
Without special mention, we employed three hidden layers for MLP; for example, if the size of predictive factors is 8, then the architecture of the neural CF layers is $32\rightarrow 16\rightarrow 8$, and the embedding size is $16$. For the NeuMF with pre-training, $\alpha$ was set to 0.5, allowing the pre-trained GMF and MLP to contribute equally to NeuMF's initialization. 

\subsection{Performance Comparison (RQ1)}

\begin{figure*}[t]
	\centering
	\begin{subfigure}[b]{0.28\textwidth}
		\centering
		\includegraphics[width=\textwidth]{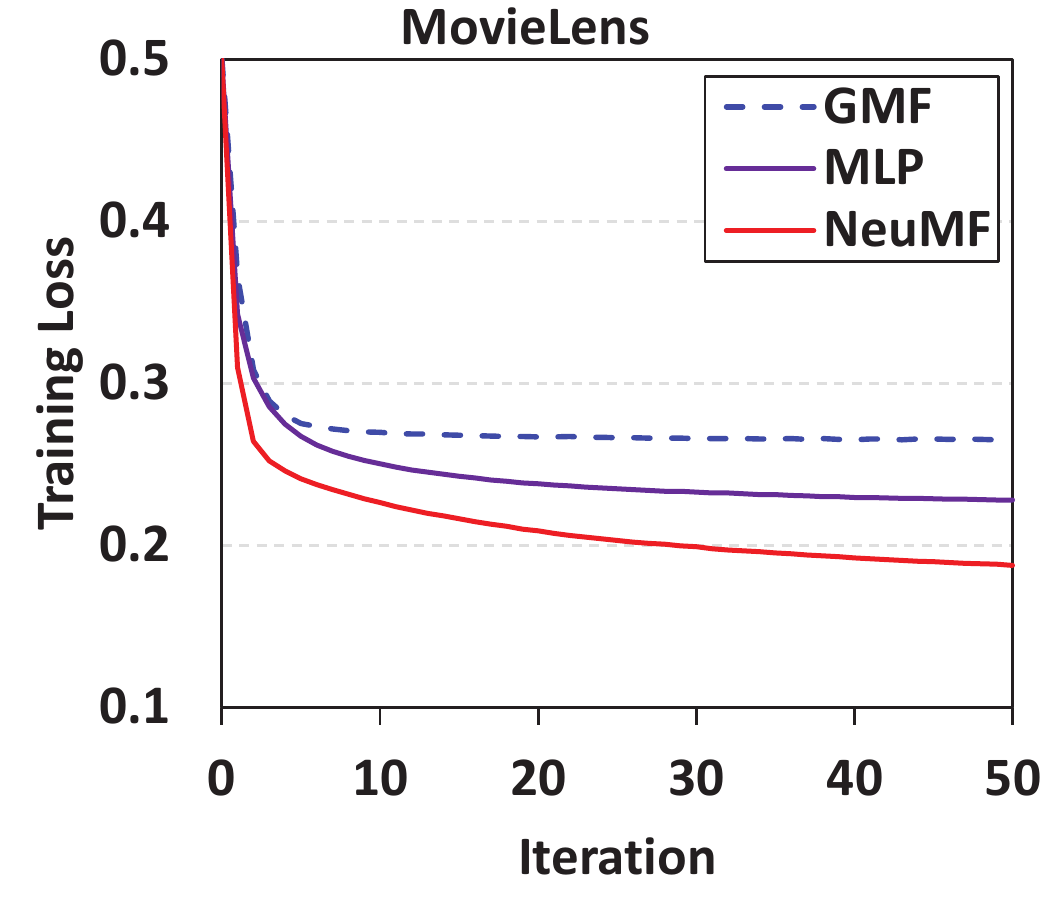}\hspace{+15pt}
		\vspace{-15pt}
		\caption{Training Loss}
		\label{fig:ml-iterations-loss}
	\end{subfigure} 
	\begin{subfigure}[b]{0.28\textwidth}
		\centering
		\includegraphics[width=\textwidth]{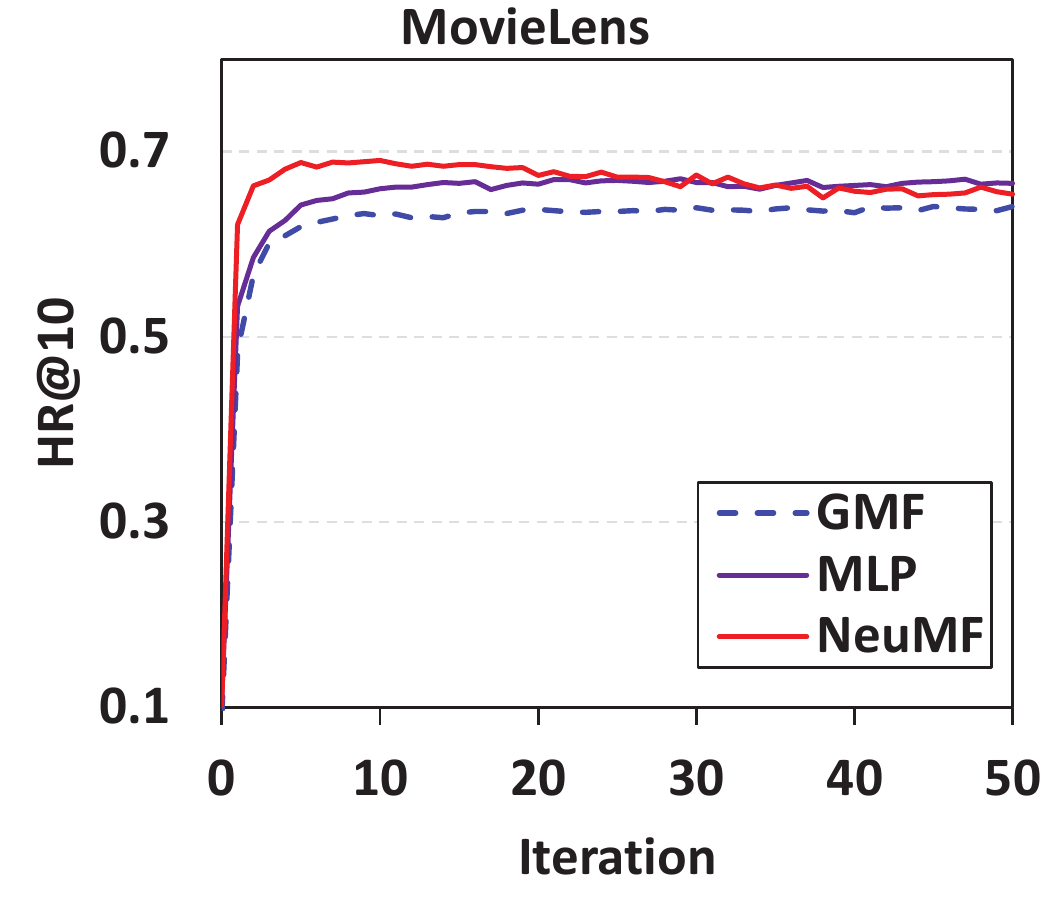}\hspace{+15pt}
		\vspace{-15pt}
		\caption{HR@10}
		\label{fig:ml-iterations-hr}
	\end{subfigure}
	\begin{subfigure}[b]{0.28\textwidth}
		\centering
		\includegraphics[width=\textwidth]{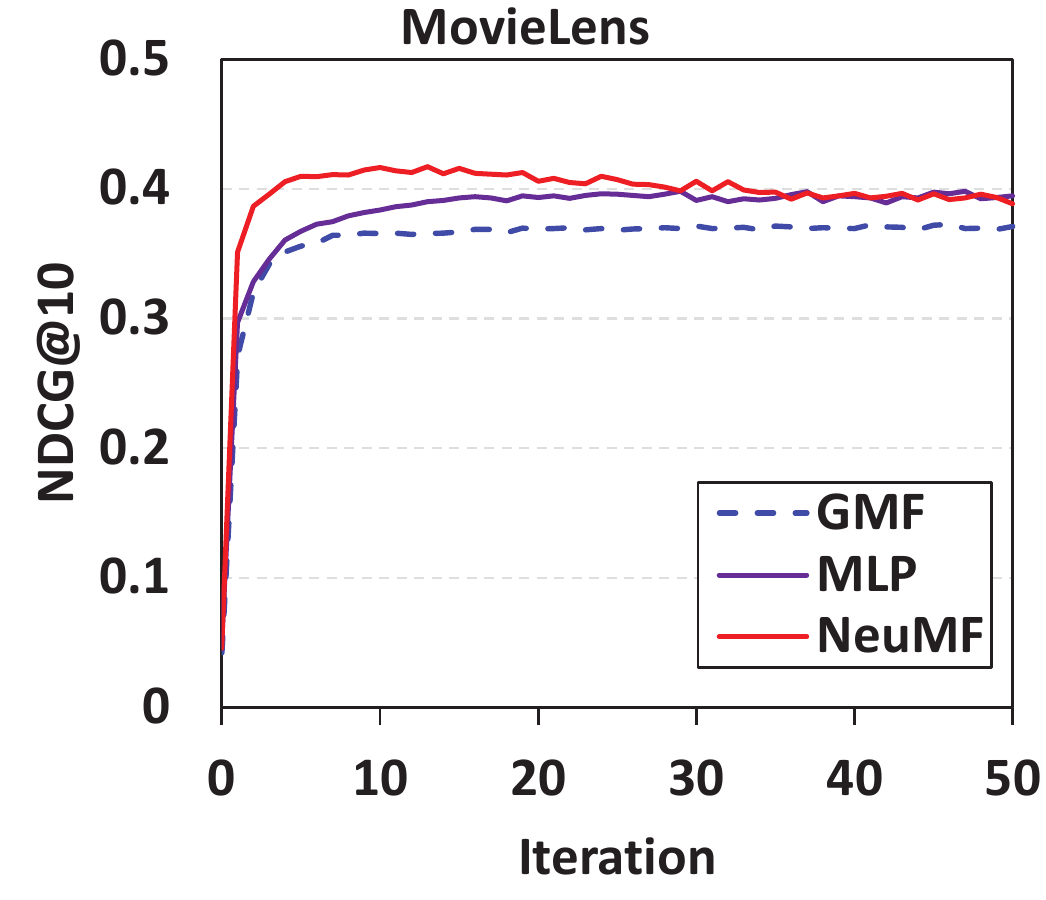}\hspace{+15pt}
		\vspace{-15pt}
		\caption{NDCG@10}
		\label{fig:ml-iterations-ndcg}
	\end{subfigure}
	\vspace{-8pt}
	\caption{Training loss and recommendation performance of NCF methods \wrt the number of iterations on MovieLens~(factors=8).}
	\vspace{-5pt}
	\label{fig:ml-iterations}
\end{figure*}

\begin{figure*}[t]
	\centering
	\begin{subfigure}[b]{0.25\textwidth}
		\centering
		\includegraphics[width=\textwidth]{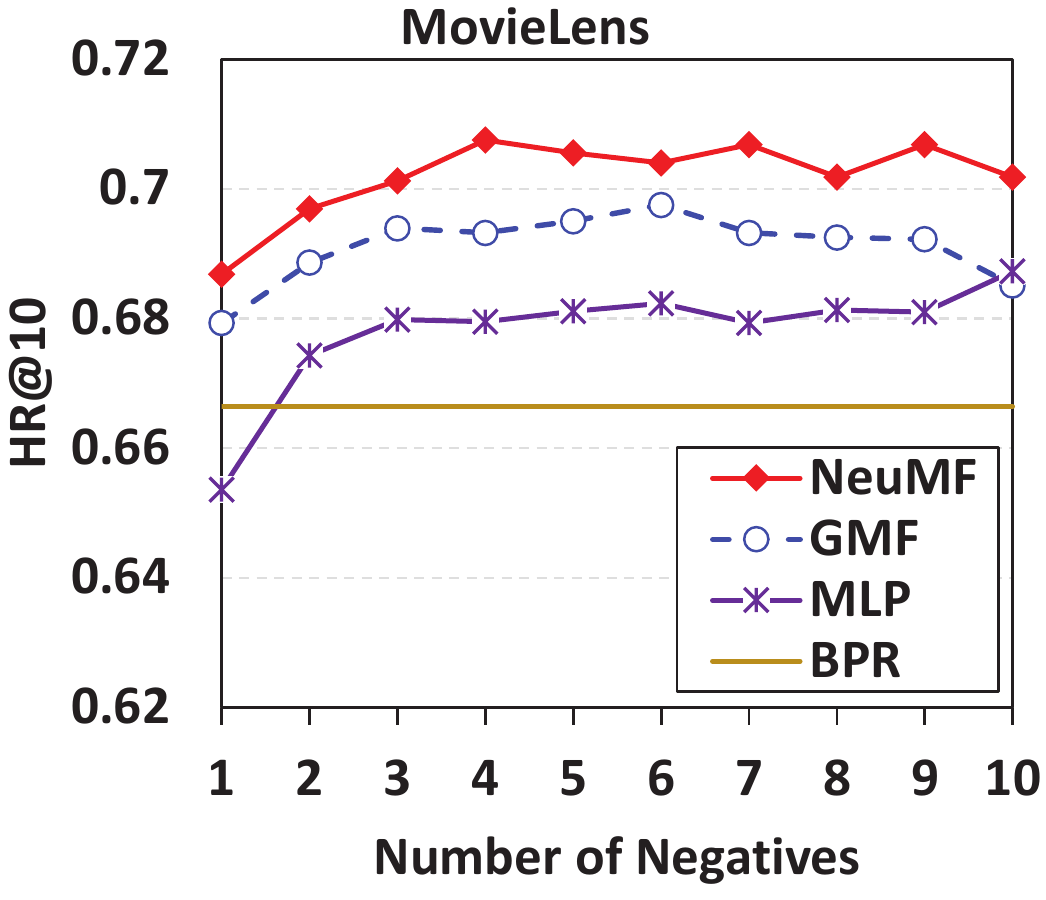}
		\vspace{-15pt}
		\caption{MovieLens --- HR@10}
		\label{fig:ml-hr-neg}
	\end{subfigure} \hspace{-7pt}
	\begin{subfigure}[b]{0.25\textwidth}
		\centering
		\includegraphics[width=\textwidth]{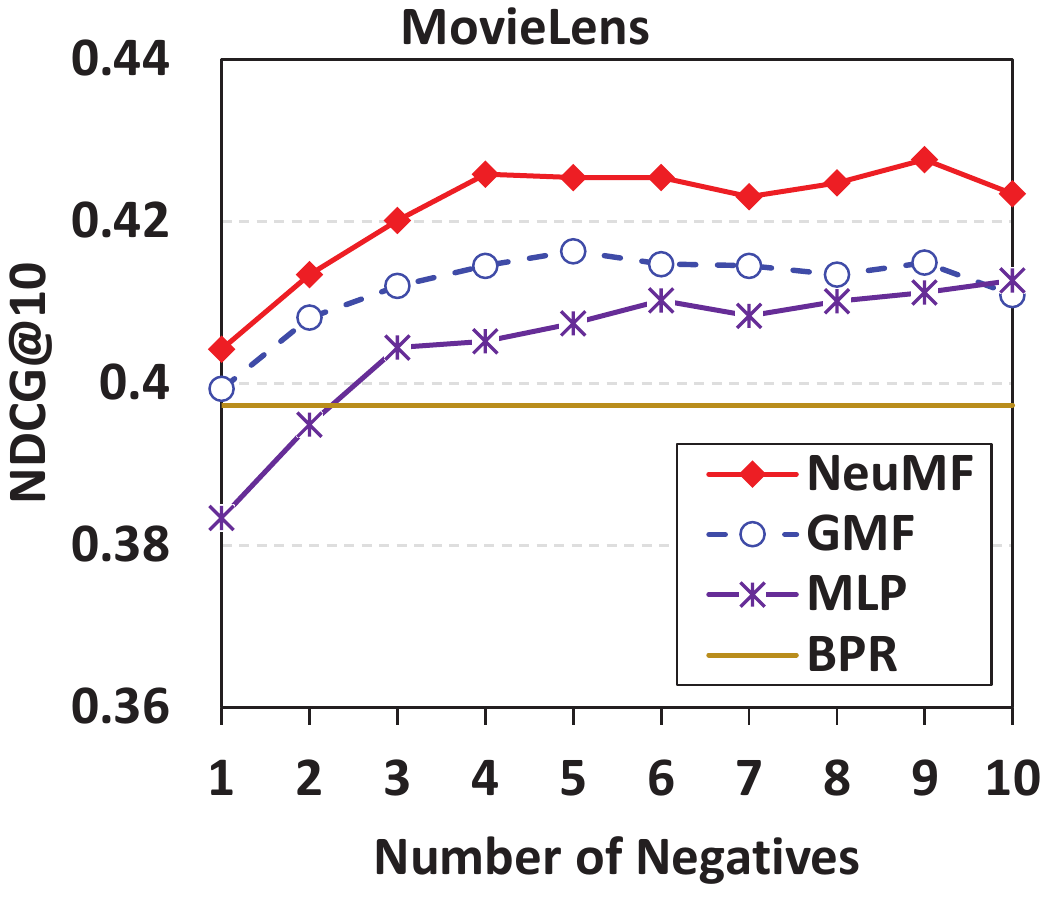}
		\vspace{-15pt}
		\caption{MovieLens --- NDCG@10}
		\label{fig:ml-ndcg-neg}
	\end{subfigure} \hspace{-7pt}
	\begin{subfigure}[b]{0.25\textwidth}
		\centering
		\includegraphics[width=\textwidth]{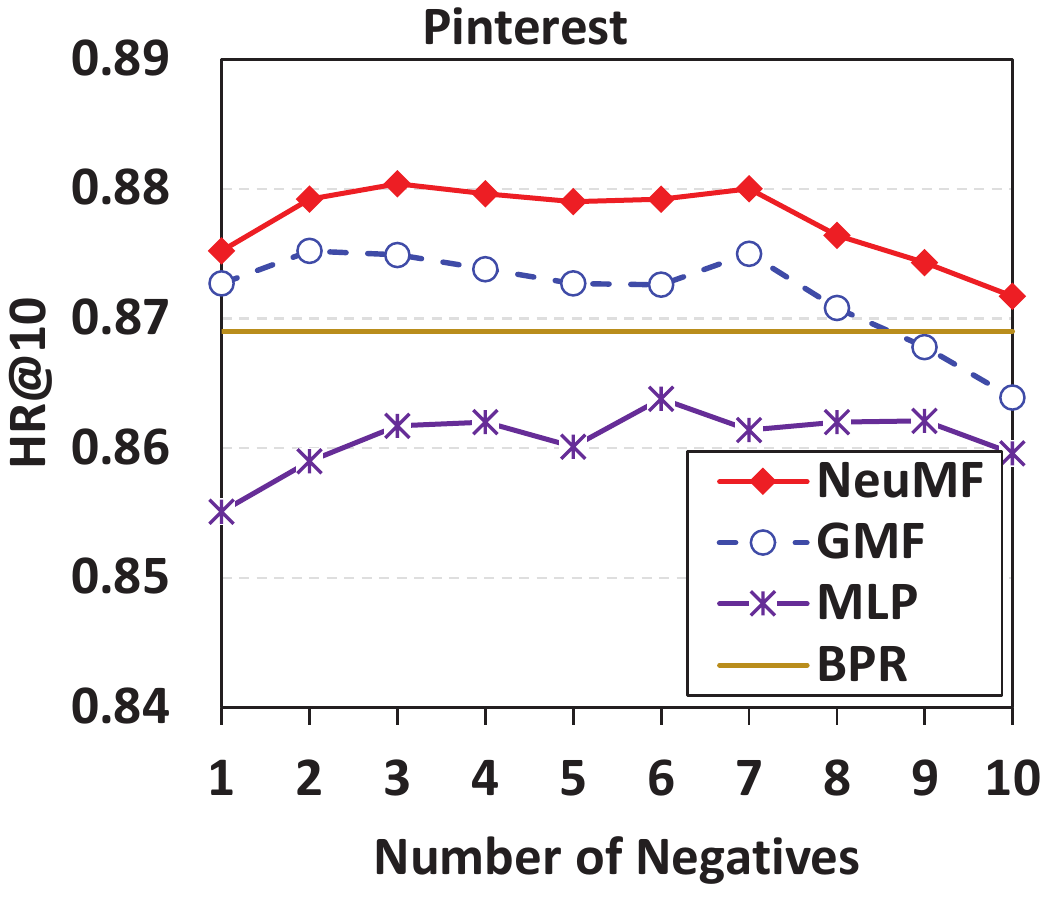}
		\vspace{-15pt}
		\caption{Pinterest --- HR@10}
		\label{fig:pinterest-hr-neg}
	\end{subfigure} \hspace{-7pt}
	\begin{subfigure}[b]{0.25\textwidth}
		\centering
		\includegraphics[width=\textwidth]{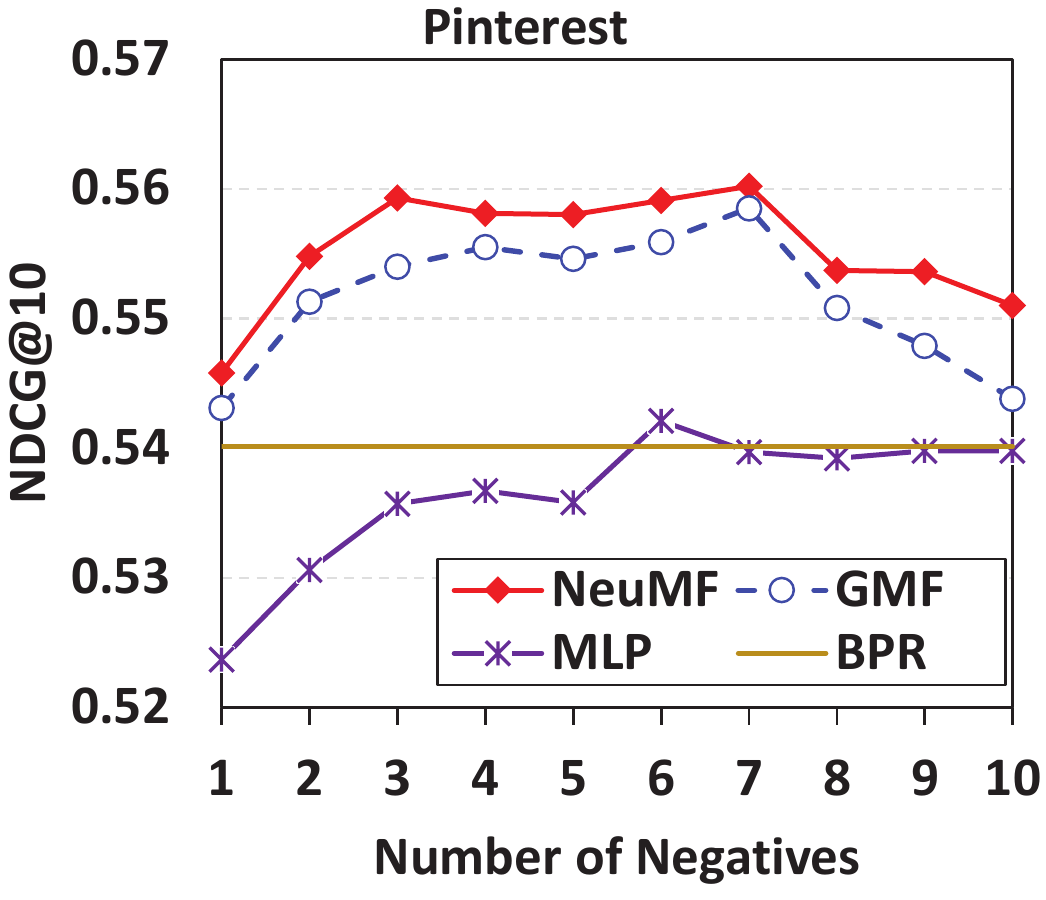}
		\vspace{-15pt}
		\caption{Pinterest --- NDCG@10}
		\label{fig:pinterest-ndcg-neg}
	\end{subfigure} \hspace{-7pt}
	\vspace{-8pt}
	\caption{Performance of NCF methods \wrt the number of negative samples per positive instance (factors=16). The performance of BPR is also shown, which samples only one negative instance to pair with a positive instance for learning.}
	\vspace{-10pt}
	\label{fig:negatives}
\end{figure*}

Figure~\ref{fig:performance-factors} shows the performance of HR@10 and NDCG@10 with respect to the number of predictive factors. For MF methods BPR and eALS, the number of predictive factors is equal to the number of latent factors. For ItemKNN, we tested different neighbor sizes and reported the best performance. Due to the weak performance of ItemPop, it is omitted in Figure~\ref{fig:performance-factors} to better highlight the performance difference of personalized methods. 

First, we can see that NeuMF achieves the best performance on both datasets, significantly outperforming the state-of-the-art methods eALS and BPR by a large margin (on average, the relative improvement over eALS and BPR is $4.5\%$ and $4.9\%$, respectively).
For Pinterest, even with a small predictive factor of 8, NeuMF substantially outperforms that of eALS and BPR with a large factor of 64. This indicates the high expressiveness of NeuMF by fusing the linear MF and non-linear MLP models. 
Second, the other two NCF methods --- GMF and MLP --- also show quite strong performance. Between them, MLP slightly underperforms GMF. Note that MLP can be further improved by adding more hidden layers (see Section~\ref{ss:exper_deep}), and here we only show the performance of three layers. 
For small predictive factors, GMF outperforms eALS on both datasets; although GMF suffers from overfitting for large factors, its best performance obtained is better than (or on par with) that of eALS. 
Lastly, GMF shows consistent improvements over BPR, admitting the effectiveness of the classification-aware log loss for the recommendation task, since GMF and BPR learn the same MF model but with different objective functions. 

Figure~\ref{fig:performance-topk} shows the performance of Top-$K$ recommended lists where the ranking position $K$ ranges from 1 to 10. 
To make the figure more clear, we show the performance of NeuMF rather than all three NCF methods. As can be seen, NeuMF demonstrates consistent improvements over other methods across positions, and we further conducted one-sample paired t-tests, verifying that all improvements are statistically significant for $p < 0.01$. For baseline methods, eALS outperforms BPR on MovieLens with about $5.1\%$ relative improvement, while underperforms BPR on Pinterest in terms of NDCG. This is consistent with \cite{fastMF}'s finding that BPR can be a strong performer for ranking performance owing to its pairwise ranking-aware learner. 
The neighbor-based ItemKNN underperforms model-based methods. 
And ItemPop performs the worst, indicating the necessity of modeling users’ personalized preferences, rather than just recommending popular items to users.

\subsubsection{Utility of Pre-training}
To demonstrate the utility of pre-training for NeuMF, we compared the performance of two versions of NeuMF --- with and without pre-training. 
For NeuMF without pre-training, we used the Adam to learn it with random initializations. 
As shown in Table~\ref{tab:pre-training}, the NeuMF with pre-training achieves better performance in most cases; only for MovieLens with a small predictive factors of 8, the pre-training method performs slightly worse. The relative improvements of the NeuMF with pre-training are $2.2\%$ and $1.1\%$ for MovieLens and Pinterest, respectively. This result justifies the usefulness of our pre-training method for initializing NeuMF. \vspace{-5pt}

\begin{table}[h]
	\begin{center}
		\caption{Performance of NeuMF with and without pre-training.}
		\small
		\vspace{-8pt}
		\label{tab:pre-training}
		\begin{tabular}{| c | c | c | c | c |} \hline
			& \multicolumn{2}{c|}{\textbf{With Pre-training}} & \multicolumn{2}{c|}{\textbf{Without Pre-training}} \\ \hline 
			\textbf{Factors} & \textbf{HR@10} & \textbf{NDCG@10} & \textbf{HR@10} & \textbf{NDCG@10} \\ \hline\hline
			\multicolumn{5}{|c|}{\textbf{MovieLens}} \\ \hline
			\textbf{8} 	& 0.684 & 0.403 & \textbf{0.688} & \textbf{0.410} \\ \hline
			\textbf{16} & \textbf{0.707}& \textbf{0.426} & 0.696 & 0.420 \\ \hline
			\textbf{32}	& \textbf{0.726} & \textbf{0.445} & 0.701 & 0.425 \\ \hline
			\textbf{64}	& \textbf{0.730} & \textbf{0.447} & 0.705 & 0.426 \\ \hline\hline
			\multicolumn{5}{|c|}{\textbf{Pinterest}} \\ \hline
			\textbf{8} 	& \textbf{0.878} & 	\textbf{0.555} &	0.869 &	0.546 \\ \hline
			\textbf{16} & \textbf{0.880} &	\textbf{0.558} &	0.871 & 0.547 \\ \hline
			\textbf{32}	& \textbf{0.879} &	\textbf{0.555} &	0.870 &	0.549 \\ \hline
			\textbf{64}	& \textbf{0.877} &	\textbf{0.552} &	0.872 &	0.551 \\ \hline
		\end{tabular}
	\end{center}
	\vspace{-15pt}
\end{table}

\vfill\eject
\subsection{Log Loss with Negative Sampling (RQ2)}
\label{ss:logloss}
To deal with the one-class nature of implicit feedback, we cast recommendation as a binary classification task. 
By viewing NCF as a probabilistic model, we optimized it with the log loss. 
Figure~\ref{fig:ml-iterations} shows the training loss (averaged over all instances) and recommendation performance of NCF methods of each iteration on MovieLens. Results on Pinterest show the same trend and thus they are omitted due to space limitation. 
First, we can see that with more iterations, the training loss of NCF models gradually decreases and the recommendation performance is improved.
The most effective updates are occurred in the first 10 iterations, and more iterations may overfit a model (\eg although the training loss of NeuMF keeps decreasing after 10 iterations, its recommendation performance actually degrades). 
Second, among the three NCF methods, NeuMF achieves the lowest training loss, followed by MLP, and then GMF. The recommendation performance also shows the same trend that NeuMF > MLP > GMF. 
The above findings provide empirical evidence for the rationality and effectiveness of optimizing the log loss for learning from implicit data. 

An advantage of pointwise log loss over pairwise objective functions~\cite{BPR,NTN} is the flexible sampling ratio for negative instances. 
While pairwise objective functions can pair only one sampled negative instance with a positive instance, we can flexibly control the sampling ratio of a pointwise loss.
To illustrate the impact of negative sampling for NCF methods, we show the performance of NCF methods \wrt different negative sampling ratios in Figure~\ref{fig:negatives}. It can be clearly seen that just one negative sample per positive instance is insufficient to achieve optimal performance, and sampling more negative instances is beneficial.
Comparing GMF to BPR, we can see the performance of GMF with a sampling ratio of one is on par with BPR, while GMF significantly betters BPR with larger sampling ratios. 
This shows the advantage of pointwise log loss over the pairwise BPR loss. 
For both datasets, the optimal sampling ratio is around 3 to 6. On Pinterest, we find that when the sampling ratio is larger than $7$, the performance of NCF methods starts to drop. It reveals that setting the sampling ratio too aggressively may adversely hurt the performance. \vspace{-5pt}

\subsection{Is Deep Learning Helpful? (RQ3)}
\label{ss:exper_deep}
As there is little work on learning user--item interaction function with neural networks, it is curious to see whether using a deep network structure is beneficial to the recommendation task. Towards this end, we further investigated MLP with different number of hidden layers. The results are summarized in Table~\ref{tab:mlp_hr} and \ref{tab:mlp_ndcg}. The MLP-3 indicates the MLP method with three hidden layers (besides the embedding layer), and similar notations for others. As we can see, even for models with the same capability, stacking more layers are beneficial to performance.
This result is highly encouraging, indicating the effectiveness of using deep models for collaborative recommendation. 
We attribute the improvement to the high non-linearities brought by stacking more non-linear layers. To verify this, we further tried stacking linear layers, using an identity function as the activation function. The performance is much worse than using the ReLU unit. 

For MLP-0 that has no hidden layers~(\ie the embedding layer is directly projected to predictions), the performance is very weak and is not better than the non-personalized ItemPop. This verifies our argument in Section~\ref{ss:mlp} that simply concatenating user and item latent vectors is insufficient for modelling their feature interactions, and thus the necessity of transforming it with hidden layers. 

\begin{table}[t]
	\begin{center}
		\caption{HR@10 of MLP with different layers.}
		\small
		\vspace{-8pt}
		\label{tab:mlp_hr}
		\begin{tabular}{| c | c | c | c | c | c |} \hline
			\textbf{Factors} & \textbf{MLP-0} & \textbf{MLP-1} & \textbf{MLP-2} & \textbf{MLP-3} & \textbf{MLP-4} \\ \hline\hline
			\multicolumn{6}{|c|}{\textbf{MovieLens}} \\ \hline
			\textbf{8} 	& 0.452 & 0.628  & 0.655  &	0.671 & \textbf{0.678} \\ \hline
			\textbf{16} & 0.454 & 0.663  & 0.674  &	0.684 &	\textbf{0.690} \\ \hline
			\textbf{32}	& 0.453 & 0.682  & 0.687  & 0.692 &	\textbf{0.699} \\ \hline
			\textbf{64}	& 0.453 & 0.687	 & 0.696  & 0.702 &	\textbf{0.707} \\ \hline\hline
			\multicolumn{6}{|c|}{\textbf{Pinterest}} \\ \hline
			\textbf{8} 	& 0.275 & 0.848 &	0.855 &	0.859 &	\textbf{0.862} \\ \hline
			\textbf{16} & 0.274 & 0.855 &	0.861 &	0.865 &	\textbf{0.867} \\ \hline
			\textbf{32}	& 0.273 & 0.861 &	0.863 &	\textbf{0.868} &	0.867 \\ \hline
			\textbf{64}	& 0.274 & 0.864 &	0.867 & 0.869 & \textbf{0.873} \\ \hline
		\end{tabular}
	\end{center}
	\vspace{-12pt}
\end{table}

\begin{table}[t]
	\begin{center}
		\caption{NDCG@10 of MLP with different layers.}
		\small
		\vspace{-8pt}
		\label{tab:mlp_ndcg}
		\begin{tabular}{| c | c | c | c | c | c |} \hline
			\textbf{Factors} & \textbf{MLP-0} & \textbf{MLP-1} & \textbf{MLP-2} & \textbf{MLP-3} & \textbf{MLP-4} \\ \hline\hline
			\multicolumn{6}{|c|}{\textbf{MovieLens}} \\ \hline
			\textbf{8} 	& 0.253 & 0.359 &	0.383 &	0.399 & \textbf{0.406} \\ \hline
			\textbf{16} & 0.252 & 0.391 &	0.402 &	0.410 & \textbf{0.415} \\ \hline
			\textbf{32}	& 0.252	& 0.406 &	0.410 & \textbf{0.425} & 0.423 \\ \hline
			\textbf{64}	& 0.251	& 0.409 &	0.417 &	0.426 &	\textbf{0.432} \\ \hline\hline
			\multicolumn{6}{|c|}{\textbf{Pinterest}} \\ \hline
			\textbf{8} 	& 0.141 & 0.526 &	0.534 &	0.536 &	\textbf{0.539} \\ \hline
			\textbf{16} & 0.141	& 0.532 &	0.536 &	0.538 &	\textbf{0.544} \\ \hline
			\textbf{32}	& 0.142	& 0.537 &	0.538 &	0.542 &	\textbf{0.546} \\ \hline
			\textbf{64}	& 0.141 & 0.538	&	0.542 &	0.545 & \textbf{0.550} \\ \hline
		\end{tabular}
	\end{center}
	\vspace{-17pt}
\end{table}

%% file: 5_related.tex
\section{Related Work}
\label{sec:related}

While early literature on recommendation has largely focused on explicit feedback~\cite{Salakhutdinov:2007,ItemCF}, recent attention is increasingly shifting towards implicit data~\cite{iCD,fastMF,UserExposure16}. The collaborative filtering (CF) task with implicit feedback is usually formulated as an item recommendation problem, for which the aim is to recommend a short list of items to users. In contrast to rating prediction that has been widely solved by work on explicit feedback, addressing the item recommendation problem is more practical but challenging~\cite{iCD,TriRank}. One key insight is to model the missing data, which are always ignored by the work on explicit feedback~\cite{SVD++,Zheng:ICML16}. 
To tailor latent factor models for item recommendation with implicit feedback, early work~\cite{Hu:2008,BPR} applies a uniform weighting where two strategies have been proposed --- which either treated all missing data as negative instances~\cite{Hu:2008} or sampled negative instances from missing data~\cite{BPR}. 
Recently, He \etal~\cite{fastMF} and Liang \etal~\cite{UserExposure16} proposed dedicated models to weight missing data, and Rendle \etal~\cite{iCD} developed an implicit coordinate descent (iCD) solution for feature-based factorization models, achieving state-of-the-art performance for item recommendation. 
In the following, we discuss recommendation works that use neural networks. 





The early pioneer work by Salakhutdinov~\etal~\cite{Salakhutdinov:2007} proposed a two-layer Restricted Boltzmann Machines (RBMs) to model users' explicit ratings on items. The work was been later extended to model the ordinal nature of ratings~\cite{Truyen:2009:OBM}.
Recently, autoencoders have become a popular choice for building recommendation systems~\cite{Sedhain:WWW15,Li:CIKM15,strub2015workshop}. The idea of user-based AutoRec~\cite{Sedhain:WWW15} is to learn hidden structures that can reconstruct a user's ratings given her historical ratings as inputs. In terms of user personalization, this approach shares a similar spirit as the item--item model~\cite{ItemCF,SLIM} that represents a user as her rated items. To avoid autoencoders learning an identity function and failing to generalize to unseen data, denoising autoencoders (DAEs) have been applied to learn from intentionally corrupted inputs~\cite{Li:CIKM15,strub2015workshop}. More recently, Zheng~\etal~\cite{Zheng:ICML16} presented a neural autoregressive method for CF.
While the previous effort has lent support to the effectiveness of neural networks for addressing CF, most of them focused on explicit ratings and modelled the observed data only. As a result, they can easily fail to learn users' preference from the positive-only implicit data. 

Although some recent works~\cite{Elkahky:2015:MVDL,van2013deep,Wang:KDD15,wang2014mm, Zhang:2016:CKB} have explored deep learning models for recommendation based on implicit feedback, they primarily used DNNs for modelling auxiliary information, such as textual description of items~\cite{Wang:KDD15}, acoustic features of musics~\cite{van2013deep,wang2014mm}, cross-domain behaviors of users~\cite{Elkahky:2015:MVDL}, and the rich information in knowledge bases~\cite{Zhang:2016:CKB}. 
The features learnt by DNNs are then integrated with MF for CF. 
The work that is most relevant to our work is \cite{Wu:WSDM16}, which presents a collaborative denoising autoencoder~(CDAE) for CF with implicit feedback. In contrast to the DAE-based CF~\cite{strub2015workshop}, CDAE additionally plugs a user node to the input of autoencoders for reconstructing the user's ratings. As shown by the authors, CDAE is equivalent to the SVD++ model~\cite{SVD++} when the identity function is applied to activate the hidden layers of CDAE. This implies that although CDAE is a neural modelling approach for CF, it still applies a linear kernel (\ie inner product) to model user--item interactions. This may partially explain why using deep layers for CDAE does not improve the performance~(\cf Section~6 of \cite{Wu:WSDM16}). Distinct from CDAE, our NCF adopts a two-pathway architecture, modelling user--item interactions with a multi-layer feedforward neural network. This allows NCF to learn an arbitrary function 
from the data, being more powerful and expressive than the fixed inner product function.

Along a similar line, learning the relations of two entities has been intensively studied in literature of knowledge graphs~\cite{TransE,NTN}. Many relational machine learning methods have been devised~\cite{RML_survey}. 
The one that is most similar to our proposal is the Neural Tensor Network~(NTN)~\cite{NTN}, which uses neural networks to learn the interaction of two entities and shows strong performance. Here we focus on a different problem setting of CF. While the idea of NeuMF that combines MF with MLP is partially inspired by NTN, our NeuMF is more flexible and generic than NTN, in terms of allowing MF and MLP learning different sets of embeddings. 

More recently, Google publicized their Wide \& Deep learning approach for App recommendation~\cite{WideDeep}. The deep component similarly uses a MLP on feature embeddings, which has been reported to have strong generalization ability. While their work has focused on incorporating various features of users and items, we target at exploring DNNs for pure collaborative filtering systems. We show that DNNs are a promising choice for modelling user--item interactions, which to our knowledge has not been investigated before. 

%% file: 6_conclusion.tex
\section{Conclusion and Future Work}
\label{sec:conclusion}

In this work, we explored neural network architectures for collaborative filtering. We devised a general framework NCF and proposed three instantiations --- GMF, MLP and NeuMF --- that model user--item interactions in different ways. 
Our framework is simple and generic; it is not limited to the models presented in this paper, but is designed to serve as a guideline for developing deep learning methods for recommendation. 
This work complements the mainstream shallow models for collaborative filtering, opening up a new avenue of research possibilities for recommendation based on deep learning. 

In future, we will study pairwise learners for NCF models and extend NCF to model auxiliary information, such as user reviews~\cite{TriRank}, knowledge bases~\cite{Zhang:2016:CKB}, and temporal signals~\cite{iCD}. 
While existing personalization models have primarily focused on individuals, it is interesting to develop models for groups of users, which help the decision-making for social groups~\cite{Hong_group,xiangwang}.
Moreover, we are particularly interested in building recommender systems for multi-media items, an interesting task but has received relatively less scrutiny in the recommendation community~\cite{Chen:2016}.
Multi-media items, such as images and videos, contain much richer visual semantics~\cite{Hong_visual,Wang_hyper} that can reflect users' interest. 
To build a multi-media recommender system, we need to develop effective methods to learn from multi-view and multi-modal data~\cite{He:WWW2014,Wang_multimodal}. 
Another emerging direction is to explore the potential of recurrent neural networks and hashing methods~\cite{DiscreteCF} for providing efficient online recommendation~\cite{fastMF,iCD}. 

\vspace{+5pt}
\noindent {\large \textbf{Acknowledgement}}

\noindent 
The authors thank the anonymous reviewers for their valuable comments, which are beneficial to the authors' thoughts on recommendation systems and the revision of the paper. 

%% file: 0_main.bbl
\begin{thebibliography}{10}

\bibitem{iCD}
I.~Bayer, X.~He, B.~Kanagal, and S.~Rendle.
\newblock A generic coordinate descent framework for learning from implicit
  feedback.
\newblock In {\em WWW}, 2017.

\bibitem{TransE}
A.~Bordes, N.~Usunier, A.~Garcia-Duran, J.~Weston, and O.~Yakhnenko.
\newblock Translating embeddings for modeling multi-relational data.
\newblock In {\em NIPS}, pages 2787--2795, 2013.

\bibitem{Chen:2016}
T.~Chen, X.~He, and M.-Y. Kan.
\newblock Context-aware image tweet modelling and recommendation.
\newblock In {\em MM}, pages 1018--1027, 2016.

\bibitem{WideDeep}
H.-T. Cheng, L.~Koc, J.~Harmsen, T.~Shaked, T.~Chandra, H.~Aradhye,
  G.~Anderson, G.~Corrado, W.~Chai, M.~Ispir, et~al.
\newblock Wide \& deep learning for recommender systems.
\newblock {\em arXiv preprint arXiv:1606.07792}, 2016.

\bibitem{Collobert:2008}
R.~Collobert and J.~Weston.
\newblock A unified architecture for natural language processing: Deep neural
  networks with multitask learning.
\newblock In {\em ICML}, pages 160--167, 2008.

\bibitem{Elkahky:2015:MVDL}
A.~M. Elkahky, Y.~Song, and X.~He.
\newblock A multi-view deep learning approach for cross domain user modeling in
  recommendation systems.
\newblock In {\em WWW}, pages 278--288, 2015.

\bibitem{Erhan:2010}
D.~Erhan, Y.~Bengio, A.~Courville, P.-A. Manzagol, P.~Vincent, and S.~Bengio.
\newblock Why does unsupervised pre-training help deep learning?
\newblock {\em Journal of Machine Learning Research}, 11:625--660, 2010.

\bibitem{Geng:2015}
X.~Geng, H.~Zhang, J.~Bian, and T.-S. Chua.
\newblock Learning image and user features for recommendation in social
  networks.
\newblock In {\em ICCV}, pages 4274--4282, 2015.

\bibitem{glorot2011deep}
X.~Glorot, A.~Bordes, and Y.~Bengio.
\newblock Deep sparse rectifier neural networks.
\newblock In {\em AISTATS}, pages 315--323, 2011.

\bibitem{cvpr16best}
K.~He, X.~Zhang, S.~Ren, and J.~Sun.
\newblock Deep residual learning for image recognition.
\newblock In {\em CVPR}, 2016.

\bibitem{TriRank}
X.~He, T.~Chen, M.-Y. Kan, and X.~Chen.
\newblock {TriRank}: Review-aware explainable recommendation by modeling
  aspects.
\newblock In {\em CIKM}, pages 1661--1670, 2015.

\bibitem{HeSIGIR2014}
X.~He, M.~Gao, M.-Y. Kan, Y.~Liu, and K.~Sugiyama.
\newblock Predicting the popularity of web 2.0 items based on user comments.
\newblock In {\em SIGIR}, pages 233--242, 2014.

\bibitem{He:WWW2014}
X.~He, M.-Y. Kan, P.~Xie, and X.~Chen.
\newblock Comment-based multi-view clustering of web 2.0 items.
\newblock In {\em WWW}, pages 771--782, 2014.

\bibitem{fastMF}
X.~He, H.~Zhang, M.-Y. Kan, and T.-S. Chua.
\newblock Fast matrix factorization for online recommendation with implicit
  feedback.
\newblock In {\em SIGIR}, pages 549--558, 2016.

\bibitem{Hong_group}
R.~Hong, Z.~Hu, L.~Liu, M.~Wang, S.~Yan, and Q.~Tian.
\newblock Understanding blooming human groups in social networks.
\newblock {\em IEEE Transactions on Multimedia}, 17(11):1980--1988, 2015.

\bibitem{Hong_visual}
R.~Hong, Y.~Yang, M.~Wang, and X.~S. Hua.
\newblock Learning visual semantic relationships for efficient visual
  retrieval.
\newblock {\em IEEE Transactions on Big Data}, 1(4):152--161, 2015.

\bibitem{Hornik:1989}
K.~Hornik, M.~Stinchcombe, and H.~White.
\newblock Multilayer feedforward networks are universal approximators.
\newblock {\em Neural Networks}, 2(5):359--366, 1989.

\bibitem{Hu:2014}
L.~Hu, A.~Sun, and Y.~Liu.
\newblock Your neighbors affect your ratings: On geographical neighborhood
  influence to rating prediction.
\newblock In {\em SIGIR}, pages 345--354, 2014.

\bibitem{Hu:2008}
Y.~Hu, Y.~Koren, and C.~Volinsky.
\newblock Collaborative filtering for implicit feedback datasets.
\newblock In {\em ICDM}, pages 263--272, 2008.

\bibitem{adam}
D.~Kingma and J.~Ba.
\newblock Adam: A method for stochastic optimization.
\newblock In {\em ICLR}, pages 1--15, 2014.

\bibitem{SVD++}
Y.~Koren.
\newblock Factorization meets the neighborhood: A multifaceted collaborative
  filtering model.
\newblock In {\em KDD}, pages 426--434, 2008.

\bibitem{Li:CIKM15}
S.~Li, J.~Kawale, and Y.~Fu.
\newblock Deep collaborative filtering via marginalized denoising auto-encoder.
\newblock In {\em CIKM}, pages 811--820, 2015.

\bibitem{UserExposure16}
D.~Liang, L.~Charlin, J.~McInerney, and D.~M. Blei.
\newblock Modeling user exposure in recommendation.
\newblock In {\em WWW}, pages 951--961, 2016.

\bibitem{RML_survey}
M.~Nickel, K.~Murphy, V.~Tresp, and E.~Gabrilovich.
\newblock A review of relational machine learning for knowledge graphs.
\newblock {\em Proceedings of the IEEE}, 104:11--33, 2016.

\bibitem{SLIM}
X.~Ning and G.~Karypis.
\newblock Slim: Sparse linear methods for top-n recommender systems.
\newblock In {\em ICDM}, pages 497--506, 2011.

\bibitem{FM}
S.~Rendle.
\newblock Factorization machines.
\newblock In {\em ICDM}, pages 995--1000, 2010.

\bibitem{BPR}
S.~Rendle, C.~Freudenthaler, Z.~Gantner, and L.~Schmidt-Thieme.
\newblock Bpr: Bayesian personalized ranking from implicit feedback.
\newblock In {\em UAI}, pages 452--461, 2009.

\bibitem{fastFM}
S.~Rendle, Z.~Gantner, C.~Freudenthaler, and L.~Schmidt-Thieme.
\newblock Fast context-aware recommendations with factorization machines.
\newblock In {\em SIGIR}, pages 635--644, 2011.

\bibitem{PMF}
R.~Salakhutdinov and A.~Mnih.
\newblock Probabilistic matrix factorization.
\newblock In {\em NIPS}, pages 1--8, 2008.

\bibitem{Salakhutdinov:2007}
R.~Salakhutdinov, A.~Mnih, and G.~Hinton.
\newblock Restricted boltzmann machines for collaborative filtering.
\newblock In {\em ICDM}, pages 791--798, 2007.

\bibitem{ItemCF}
B.~Sarwar, G.~Karypis, J.~Konstan, and J.~Riedl.
\newblock Item-based collaborative filtering recommendation algorithms.
\newblock In {\em WWW}, pages 285--295, 2001.

\bibitem{Sedhain:WWW15}
S.~Sedhain, A.~K. Menon, S.~Sanner, and L.~Xie.
\newblock Autorec: Autoencoders meet collaborative filtering.
\newblock In {\em WWW}, pages 111--112, 2015.

\bibitem{NTN}
R.~Socher, D.~Chen, C.~D. Manning, and A.~Ng.
\newblock Reasoning with neural tensor networks for knowledge base completion.
\newblock In {\em NIPS}, pages 926--934, 2013.

\bibitem{srivastava2012multimodal}
N.~Srivastava and R.~R. Salakhutdinov.
\newblock Multimodal learning with deep boltzmann machines.
\newblock In {\em NIPS}, pages 2222--2230, 2012.

\bibitem{strub2015workshop}
F.~Strub and J.~Mary.
\newblock Collaborative filtering with stacked denoising autoencoders and
  sparse inputs.
\newblock In {\em NIPS Workshop on Machine Learning for eCommerce}, 2015.

\bibitem{Truyen:2009:OBM}
T.~T. Truyen, D.~Q. Phung, and S.~Venkatesh.
\newblock Ordinal boltzmann machines for collaborative filtering.
\newblock In {\em UAI}, pages 548--556, 2009.

\bibitem{van2013deep}
A.~Van~den Oord, S.~Dieleman, and B.~Schrauwen.
\newblock Deep content-based music recommendation.
\newblock In {\em NIPS}, pages 2643--2651, 2013.

\bibitem{Wang:KDD15}
H.~Wang, N.~Wang, and D.-Y. Yeung.
\newblock Collaborative deep learning for recommender systems.
\newblock In {\em KDD}, pages 1235--1244, 2015.

\bibitem{Wang_anchor}
M.~Wang, W.~Fu, S.~Hao, D.~Tao, and X.~Wu.
\newblock Scalable semi-supervised learning by efficient anchor graph
  regularization.
\newblock {\em IEEE Transactions on Knowledge and Data Engineering},
  28(7):1864--1877, 2016.

\bibitem{Wang_multimodal}
M.~Wang, H.~Li, D.~Tao, K.~Lu, and X.~Wu.
\newblock Multimodal graph-based reranking for web image search.
\newblock {\em IEEE Transactions on Image Processing}, 21(11):4649--4661, 2012.

\bibitem{Wang_hyper}
M.~Wang, X.~Liu, and X.~Wu.
\newblock Visual classification by l1 hypergraph modeling.
\newblock {\em IEEE Transactions on Knowledge and Data Engineering},
  27(9):2564--2574, 2015.

\bibitem{xiangwang}
X.~Wang, L.~Nie, X.~Song, D.~Zhang, and T.-S. Chua.
\newblock Unifying virtual and physical worlds: Learning towards local and
  global consistency.
\newblock {\em ACM Transactions on Information Systems}, 2017.

\bibitem{wang2014mm}
X.~Wang and Y.~Wang.
\newblock Improving content-based and hybrid music recommendation using deep
  learning.
\newblock In {\em MM}, pages 627--636, 2014.

\bibitem{Wu:WSDM16}
Y.~Wu, C.~DuBois, A.~X. Zheng, and M.~Ester.
\newblock Collaborative denoising auto-encoders for top-n recommender systems.
\newblock In {\em WSDM}, pages 153--162, 2016.

\bibitem{Zhang:2016:CKB}
F.~Zhang, N.~J. Yuan, D.~Lian, X.~Xie, and W.-Y. Ma.
\newblock Collaborative knowledge base embedding for recommender systems.
\newblock In {\em KDD}, pages 353--362, 2016.

\bibitem{DiscreteCF}
H.~Zhang, F.~Shen, W.~Liu, X.~He, H.~Luan, and T.-S. Chua.
\newblock Discrete collaborative filtering.
\newblock In {\em SIGIR}, pages 325--334, 2016.

\bibitem{zhang2014start}
H.~Zhang, Y.~Yang, H.~Luan, S.~Yang, and T.-S. Chua.
\newblock Start from scratch: Towards automatically identifying, modeling, and
  naming visual attributes.
\newblock In {\em MM}, pages 187--196, 2014.

\bibitem{Zheng:ICML16}
Y.~Zheng, B.~Tang, W.~Ding, and H.~Zhou.
\newblock A neural autoregressive approach to collaborative filtering.
\newblock In {\em ICML}, pages 764--773, 2016.

\end{thebibliography}
